\documentclass[a4paper,11pt]{article}

\usepackage{jinstpub}

\usepackage{siunitx}

\newcommand{\znbb}{\textrm{0}\nu\beta\beta}

\title{The EXO-200 detector, part II: Auxiliary systems}

\author[a]{N.~Ackerman,}
\author[b, 1]{J.~Albert,}
\note{Now at Levatas, Inc., Palm Beach Gardens, FL, USA}
\author[c, 2]{M.~Auger,}
\note{Now at Tofwerk, Thun, Switzerland}
\author[d, 3]{D.J.~Auty,}
\note{Now at University of Alberta, Edmonton, AB, Canada}
\author[e, 4]{I.~Badhrees,}
\note{Permanent position with  King Abdulaziz City for Science and Technology, Riyadh, Saudi Arabia}
\author[f]{P.S.~Barbeau,}
\author[g]{L.~Bartoszek,}
\author[c, 5]{E.~Baussan,}
\note{Now at Universite de Strasbourg, CNRS, IPHC UMR 7178,F-67000 Strasbourg, France}
\author[h]{V.~Belov,}
\author[i, 6]{C.~Benitez-Medina,}
\note{Now at Intel, Hillsboro OR 97124, USA}
% ...
\affiliation[a]{SLAC National Accelerator Laboratory, Menlo Park, CA 94025, USA}
\affiliation[b]{Department of Physics and CEEM, Indiana University, Bloomington, IN 47405, USA}
\affiliation[c]{LHEP, Albert Einstein Center, University of Bern, Bern CH-3012, Switzerland}
\affiliation[d]{Department of Physics and Astronomy, University of Alabama, Tuscaloosa, AL 35487, USA}
\affiliation[e]{Department of Physics, Carleton University, Ottawa, ON K1S 5B6, Canada}
\affiliation[f]{Department of Physics, Duke University, and Triangle Universities Nuclear Laboratory (TUNL), Durham, NC 27708, USA}
\affiliation[g]{Physics Department, Stanford University, Stanford, CA 94305, USA}
\affiliation[h]{Institute for Theoretical and Experimental Physics named by A. I. Alikhanov of National Research Center ``Kurchatov Institute'', Moscow 117218, Russia}
\affiliation[i]{Physics Department, Colorado State University, Fort Collins, CO 80523, USA}
% ...

\abstract{The EXO-200 experiment searched for neutrinoless double-beta decay of $^{136}$Xe with a single-phase liquid xenon detector.  It used an active mass of \SI{110}{kg} of 80.6\%-enriched liquid xenon in an ultra-low background time projection chamber with ionization and scintillation detection and readout.  This paper describes the design and performance of the various support systems necessary for detector operation, including cryogenics, xenon handling, and controls.  Novel features of the system were driven by the need to protect the thin-walled detector chamber containing the liquid xenon, to achieve high chemical purity of the Xe, and to maintain thermal uniformity across the detector.}


\begin{document}

\maketitle

\section{Introduction}
\label{intro}

The observation of neutrinoless double-beta decay ($0\nu\beta\beta$) would prove that neutrinos are Majorana fermions and be the first evidence of the non-conservation of lepton number.  The EXO-200 experiment, located in the Waste Isolation Pilot Plant (WIPP) near Carlsbad, NM \cite{wipp}, was one of the leading experiments in the search for the decay and the first using a kilo-mole of decaying isotope.  The EXO-200 liquid xenon (LXe) time projection chamber (TPC) has been described in the first of a series of papers \cite{Auger:2012gs} detailing the design, operation and performance of the EXO-200 apparatus.  In this second installment, the various cryogenic, xenon handling, and control systems are discussed, and relevant performance data are provided.  The goals of these ancillary systems were to provide a safe, stable, thermally uniform, and reliable cryogenic environment for the TPC, to maintain the very high chemical purity in the LXe necessary for satisfactory TPC performance, and to maintain overall ultra-low radioactivity, all in an experiment deployed deep underground. 

Adding to the challenge, the thickness of the copper vessel that contained the LXe and the TPC was minimized to reduce radioactive backgrounds \cite{Auger:2012gs}, resulting in severe differential pressure (dP) constraints to protect the instrument and the expensive supply of isotopically enriched LXe.  The TPC vessel was designed to reliably support a maximum \SI{35}{kPa} differential pressure in either direction, and the fairly elaborate EXO-200 control system was designed to maintain dP to within about \SI{4}{kPa} during normal operation, experimental start-up and shut-down procedures, power outages, and other extraordinary circumstances.

\begin{figure}
	\centering
	\includegraphics[width=5in]{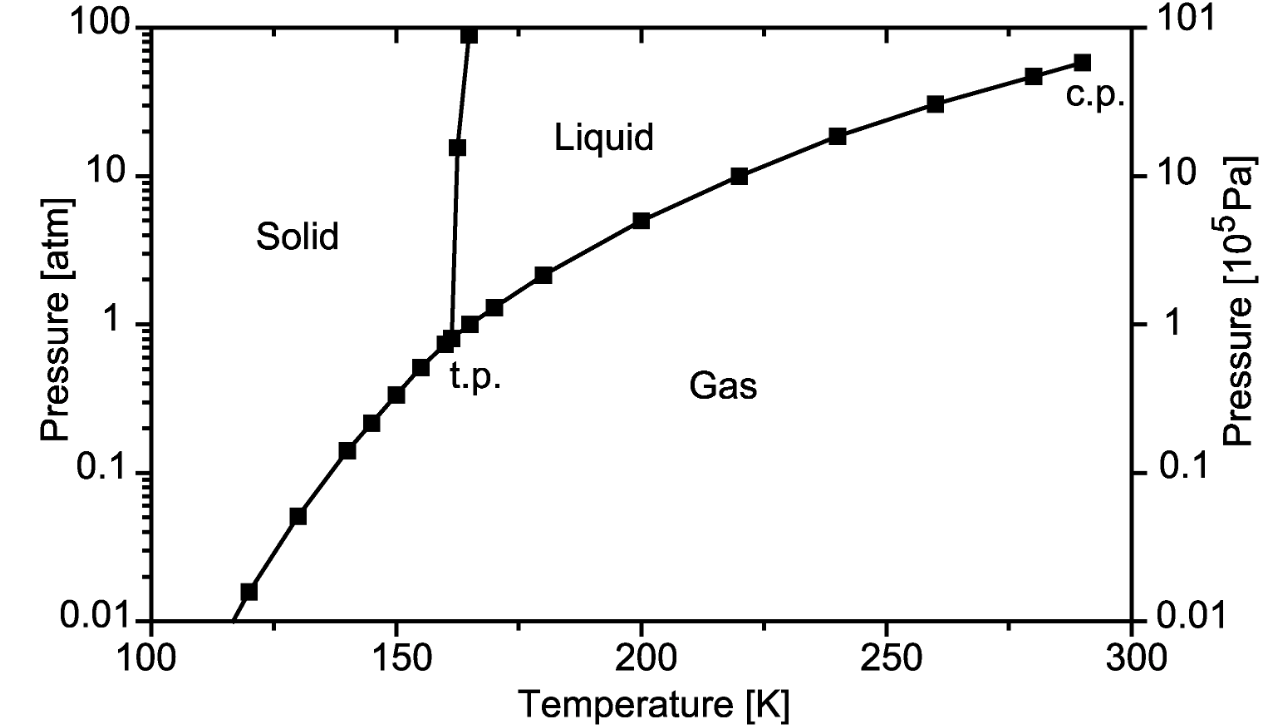}
	\caption{Xe phase diagram.  Figure taken from \cite{xe_phase_ref}. \label{fig:xe_phase}}
\end{figure}

The target parameters for temperature stability were fundamentally constrained by the properties of LXe and, more significantly, by the choice of large area avalanche photo diodes (LAAPDs) for scintillation light detection in the TPC.  Xenon at atmospheric pressure (\SI{101.3}{kPa}) liquifies at \SI{165.1}{K} and freezes at 161.4 K  \cite{nist_thermophysical}.  A phase diagram for Xe is shown in Figure \ref{fig:xe_phase}.  In addition, the LAAPDs exhibit strong temperature dependent gain, measured to be about \SI[per-mode=symbol]{5}{\percent\per\K} near 170 K \cite{Neilson:2009kf}.  The design criteria were set at temporal stability of \SI{0.1}{K}, as well as spatial uniformity across the two vertically oriented LAAPD instrumented arrays in the TPC (roughly \SI{40}{cm} diameter) of \SI{0.1}{K}, in order to ensure that temperature dependent gain-variation effects were not a limiting factor for energy resolution.

The requirements on xenon purity were less sharply defined, but were roughly targeted to keep electron capture corrections to the ionization signal from adding significantly to the detector resolution.  A fraction $1 - e^{-t/\tau}$ of ionized charge drifting in LXe is lost after time t, where $\tau$ is the electron lifetime.  The electron lifetime is therefore defined to be the drift time at which 63\% of the charge has been lost.  Assuming that corrections to the measured energy in the ionization channel can be determined to 10\% relative uncertainty, a maximum total drift time of \SI{115}{\us} implied that the electron lifetime should be greater than \SI{1}{ms} to limit the impact on the energy resolution to less than 1\%.  In R\&D studies electron lifetimes better than this were obtainable, but it was known that such high electron lifetimes (and the $< 1$ ppb oxygen-equivalent concentrations that they imply) generally require that the xenon be recirculated continuously through a purification system.  A design maximum recirculation rate of \SI{20}{SLPM} was established based on the specifications of the particular purifier selected for use.

In what follows, relevant EXO-200 systems are described, the extent to which the EXO-200 system design was successful in reaching expectations is discussed, and lessons learned along the way are detailed.  The auxiliary systems were commissioned at WIPP between October and December 2009 through an exercise in which a stainless steel ``dummy vessel'' was filled with natural LXe and recirculation through the purifiers was demonstrated, all while maintaining control of dP.  An engineering run with natural xenon was carried out December 2010 - January 2011 following the installation of the TPC, which was then filled with enriched xenon in early 2011.  Phase I of detector operation began in May 2011 and ended when events underground at WIPP forced the facility's temporary closure in February 2014.  After recovering and upgrading the experiment, Phase II ran from May 2016 to December 2018 \cite{phaseii_prl, full_dataset_prl}.  The detector was decommissioned in January 2019.  Results for the $^{136}$Xe $0\nu\beta\beta$ decay search using the complete dataset were reported in \cite{full_dataset_prl}.

\section{Cryogenic System}

Nearly all of the fundamental features of the EXO-200 cryogenic system, which includes the cryostat, refrigeration scheme, heat transfer medium, and controls, were prototyped in a small-scale setup at the SLAC National Accelerator Laboratory.  This apparatus included a \SI{0.6}{\liter} liquid xenon purity monitor (the proxy for a TPC) enclosed in a \SI{30}{\liter} dewar filled with a low temperature heat transfer fluid and cooled by a closed-cycle Polycold PFC-550 refrigerator \cite{polycold} via an immersed refrigerant coil.  Temperature control was "on/off" style with a solenoid valve in the refrigerator, via LabVIEW \cite{Labview} software.

\subsection{Heat Transfer Fluid}

The xenon vessel was cooled through a heat transfer fluid (HTF), rather than directly, in order to maximize the thermal uniformity at the TPC, to provide thermal inertia to the system, and to provide additional shielding from radioactive decays in the cryostat and from other external sources.  Early Monte Carlo studies determined that the EXO-200 TPC vessel should be shielded by no less than 50 cm of HTF in all directions.  Such a large total mass of HTF in close proximity to the xenon vessel meant that its radiopurity requirements were extremely high.  It was expected that this requirement would be satisfied by the class of fluorocarbons used as low temperature HTFs. The three HTFs considered were FC-87 \cite{FC87}, HFE-7000 \cite{HFE-7000}, and HFE-7100 \cite{HFE-7100}, and the former two were tested in the prototype lab.  While FC-87 has much higher bulk resistivity, favorable from the point of view of the high-voltage design, HFE-7000 was chosen for its favorable thermal characteristics and availability.  

The properties of HFE-7000 (henceforth ``HFE'') are given in Table \ref{tab:HFEprop}.  It retains usefully low viscosity, important for convection, at operating temperatures near \SI{168}{K}, and its freezing point is comfortably below this level.  In addition, it is a hydrofluroether (CF$_3$CF$_2$CF$_2$OCH$_3$) - and three hydrogen atoms per molecule mean that it provides some slow neutron attenuation (when cold, about 25\% water equivalent by volume).  The limits on the concentrations of $^{40}$K, $^{232}$Th, and $^{238}$U obtained by neutron activation analysis were found to be among the lowest of the detector materials studied \cite{Background_NIM}.  These limits imply a small contribution to detector backgrounds \cite{Auger:2012gs}, consistent with analysis of physics data \cite{Background_PRC}.

\begin{table}
\begin{center}
\begin{tabular}{|r|r|}
\hline
Property & Value\\
\hline
Molecular Weight & \SI[per-mode=symbol]{200}{\g\per\mol}\\
Freezing point & \SI{150.65}{K}\\
Boiling point @ 1 atm & \SI{307}{K}\\
Heat of Vaporization & \SI[per-mode=symbol]{142}{\J\per\g}\\
Solubility of N$_2$ & 55 vol $\%$\\
Density vs T & (2.26 - T/347.2) \si[per-mode=symbol]{\g\per\cm^3}\\
Specific heat vs T & (382 + 3.08$\cdot$T) \si[per-mode=symbol]{\J\per{\kg\cdot\K}}\\
Thermal conductivity vs T & (0.13 - T/5102) \si[per-mode=symbol]{\W\per{\m\cdot\K}}\\
\hline
\end{tabular}
\end{center}
\caption{Relevant properties of HFE-7000 at \SI{298}{K}, unless otherwise noted.}
\label{tab:HFEprop}
\end{table}

\subsection{The Cryostat}
\label{sec:cryostat}

The central TPC vessel was immersed in a cryostat filled with HFE maintained at the temperature of LXe around \SI{168}{K}.  The cryostat consisted of two coaxial vessels, the inner (IV) and the outer (OV) vessel, separated by a vacuum insulation gap. Both vessels were made from Aurubis  \cite{aurubis} NOSV copper (UNS C11000), which was shown to be very low in radioactivity \cite{Background_NIM}. The copper was purchased in two batches of plates \SI{2}{\meter}$\times$\SI{1}{\meter}$\times$\SI{27}{\mm}. All the components were cut out of these plates, machined and bent, and welded together by SDMS, La Chaudronnerie Blanche \cite{SDMS} and subcontractors. 

\begin{figure}[htb]
\begin{center}
\includegraphics[height=8cm]{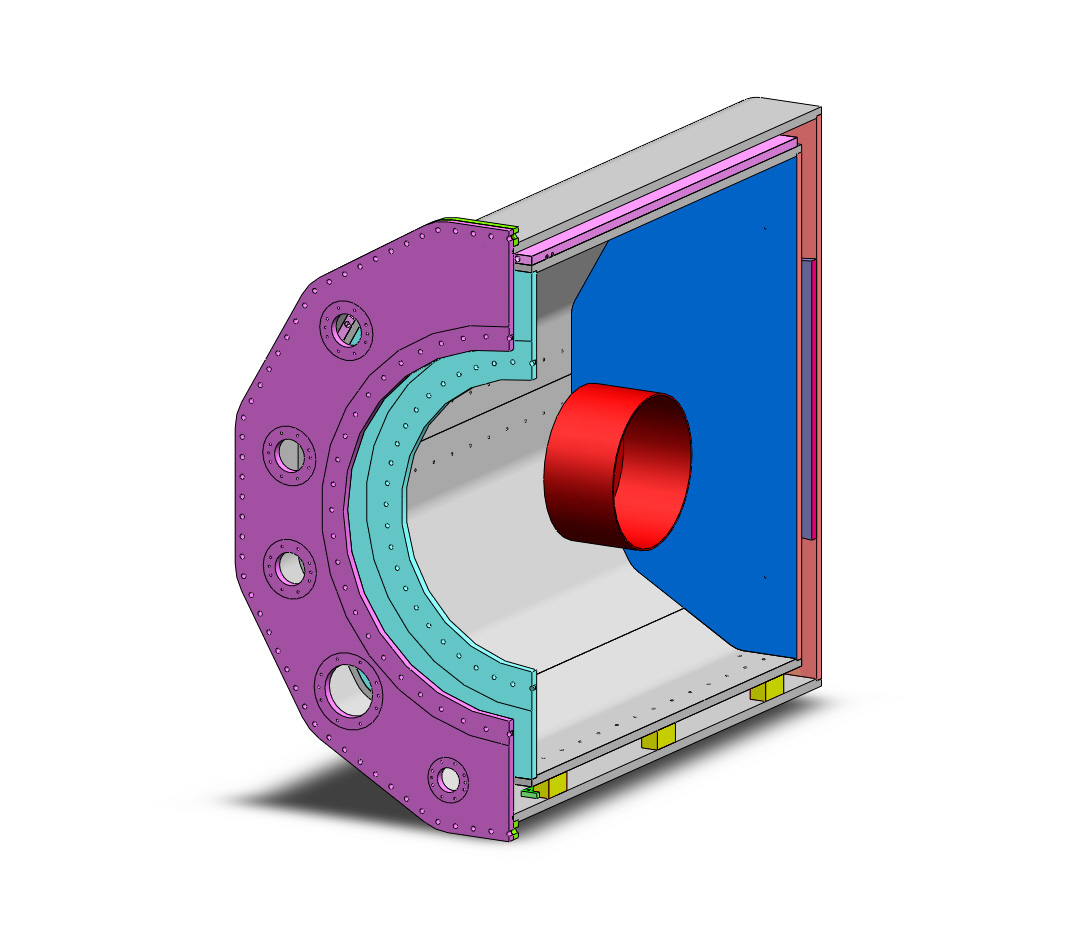}
\caption{Cutaway drawing of the EXO-200 cryostat, which consisted of the nested inner (IV) and outer (OV) vessels shown in grey.  Colors indicate the OV front flange (purple), IV front flange (blue-green), heat-exchangers (pink), IV back plate (blue), OV back plate (orange), and teflon mounting blocks (yellow).  The location of the central TPC is indicated in red.}
\label{fi:IVOV}
\end{center}
\end{figure}

The geometry of the IV and OV is shown in Figure \ref{fi:IVOV}.   Both vessels were dodecagonal prisms, a shape chosen to reduce, compared to rolling, the contact area with tooling, thus decreasing the risk of surface contamination.  The internal dimensions of the IV were \SI{1440}{\mm} in length with an apothem of \SI{720}{\mm}. For the OV these numbers were, respectively, \SI{1651}{\mm} and \SI{803}{\mm}. The lateral insulation gap was \SI{56}{\mm}. The inner vessel rested on three PTFE blocks, positioned on the bottom of the OV. The block nearest the door constrained the position of the IV, and was held in place by bars screwed in the bottom plate of the OV. Two extra blocks were placed on the lower sides at the 4 o'clock and 8 o'clock positions to give extra lateral stability.

\begin{figure}[htb]
\begin{center}
\includegraphics[height=8cm]{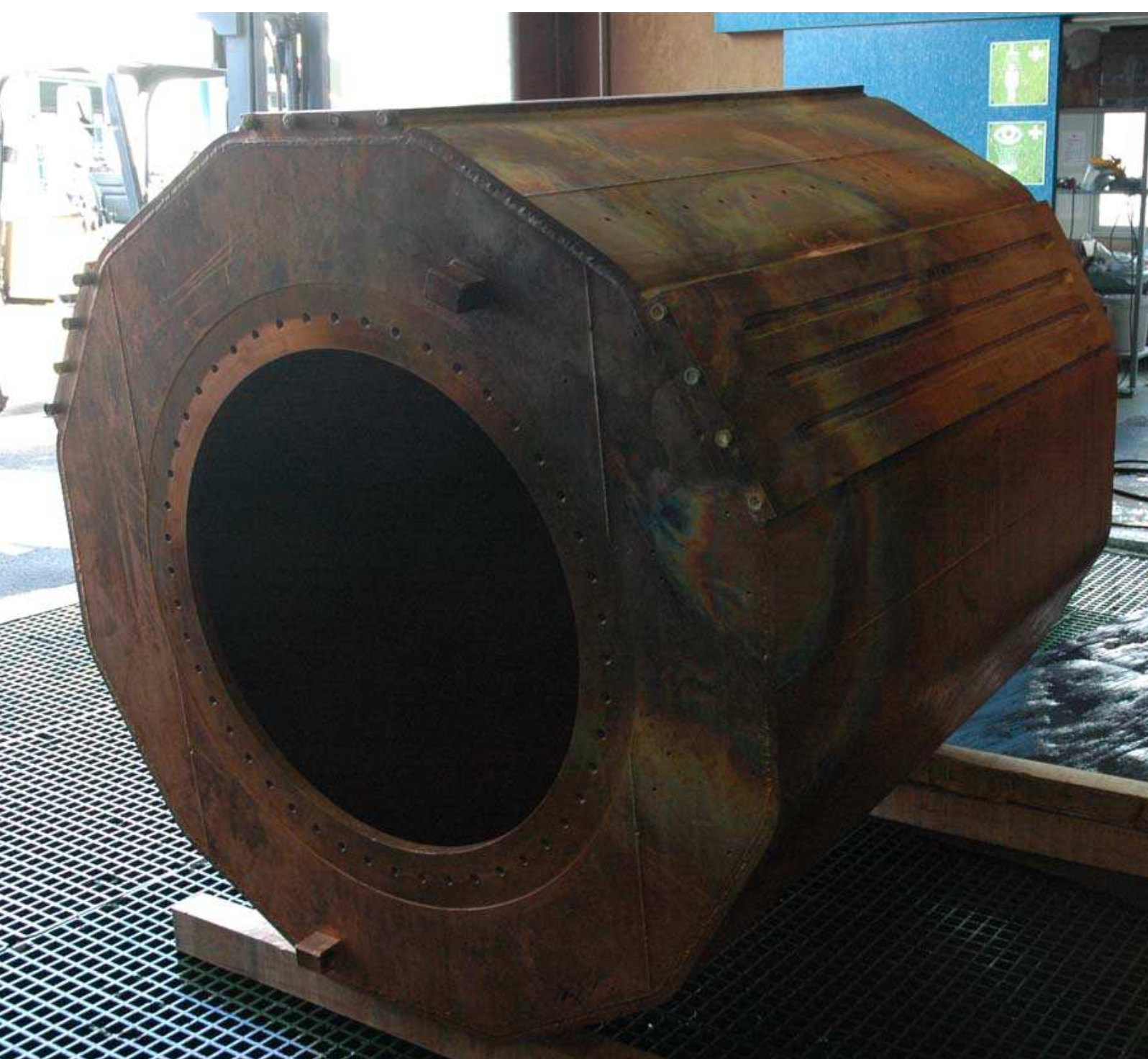}
\caption{The IV during construction at SDMS, showing the central access hole for the TPC, the two HFE ports on the front flange, and the three heat exchangers on top.  The completed vessels were subsequently cleaned as discussed in the text.} \label{fi:IV_HX}
\end{center}
\end{figure}

The IV is shown during construction in Figure \ref{fi:IV_HX}.  A front flange was welded to it with an \SI{836}{\mm} circular opening to allow for the installation of the TPC.  As described in detail in \cite{Auger:2012gs}, the TPC with all its services was cantilevered off a copper door sealed against the flange by a custom spring-loaded indium-plated phosphor bronze seal \cite{Jetseal} designed to account for the softness of the copper surfaces.  Two small ports, one at the top and the other at the bottom, were used to evacuate the vessel and to fill and to remove the HFE.  All the other feedthroughs, for LXe inlet and outlet, TPC high voltage cable, and flat cables for the wire and APD signals, were mounted on the door. 

To allow for the installation of the IV, the OV front plate was bolted onto a flange welded to the side walls.  The front plate had a \SI{1058}{\mm} diameter central access hole to allow for the installation of the IV door with the TPC.  Flanges mounted on the front plate carried feedthroughs for refrigeration lines, the source calibration guide tube, thermocouples, the HFE inlet and outlet, and evacuation.  Ports for the TPC services were similarly mounted on the door sealed over the central hole.  Except for the flat cables, the cryostat  and TPC services continued into copper vacuum jackets screwed to the front plate or its lid.  Copper conduits enclosed the flat cables on the way to the copper enclosures for the front end electronics between the front lead shielding walls, as discussed in \cite{Auger:2012gs}.  The door and all flanges mounted on the room-temperature OV were sealed with elastomer o-rings.

\begin{figure}[ht]
\begin{center}
\includegraphics[clip=true, trim=25cm 0cm 5cm 0cm, height=6cm]{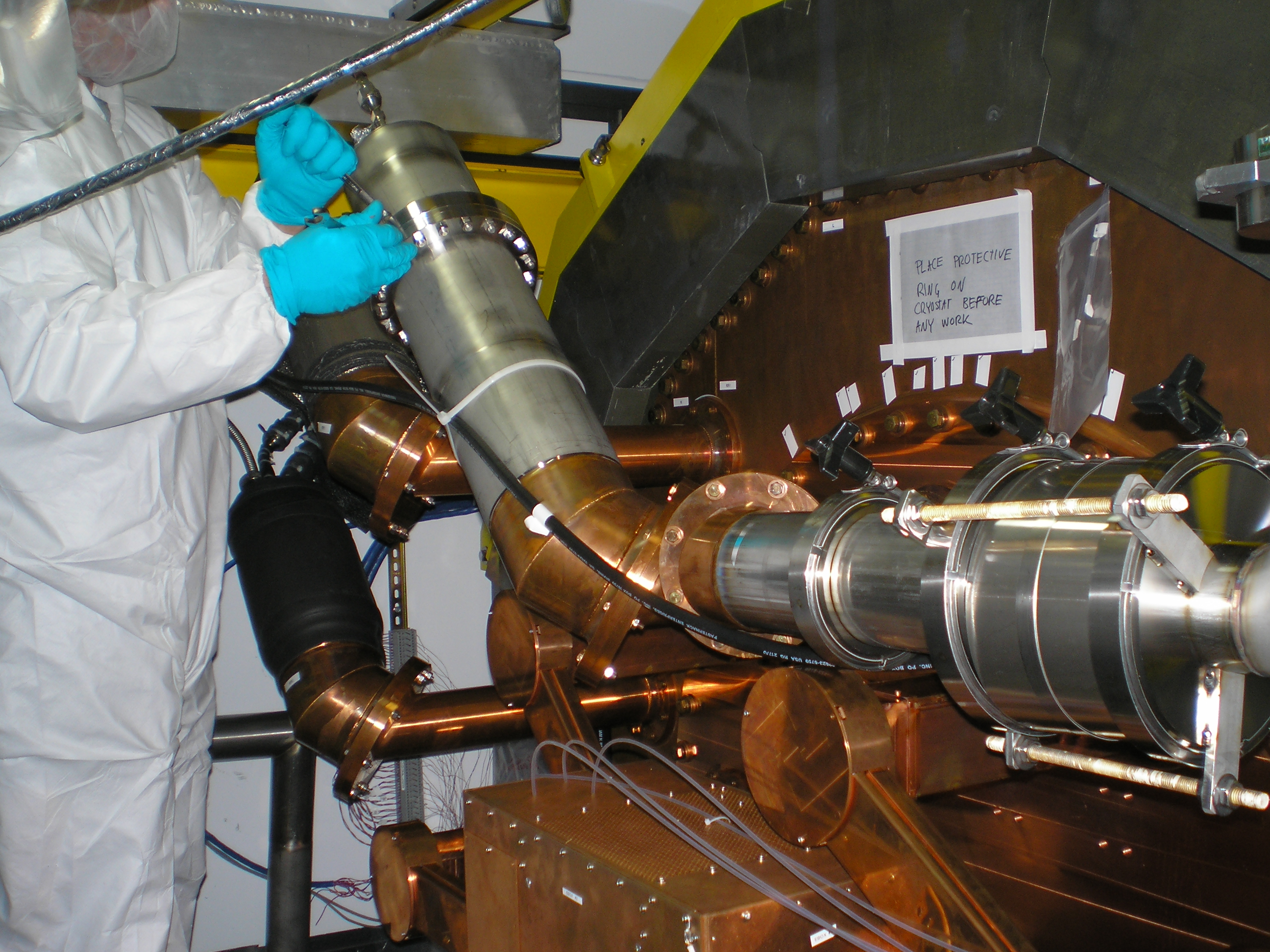}
\includegraphics[clip=true, trim=35cm 20cm 30cm 15cm, height=6cm]{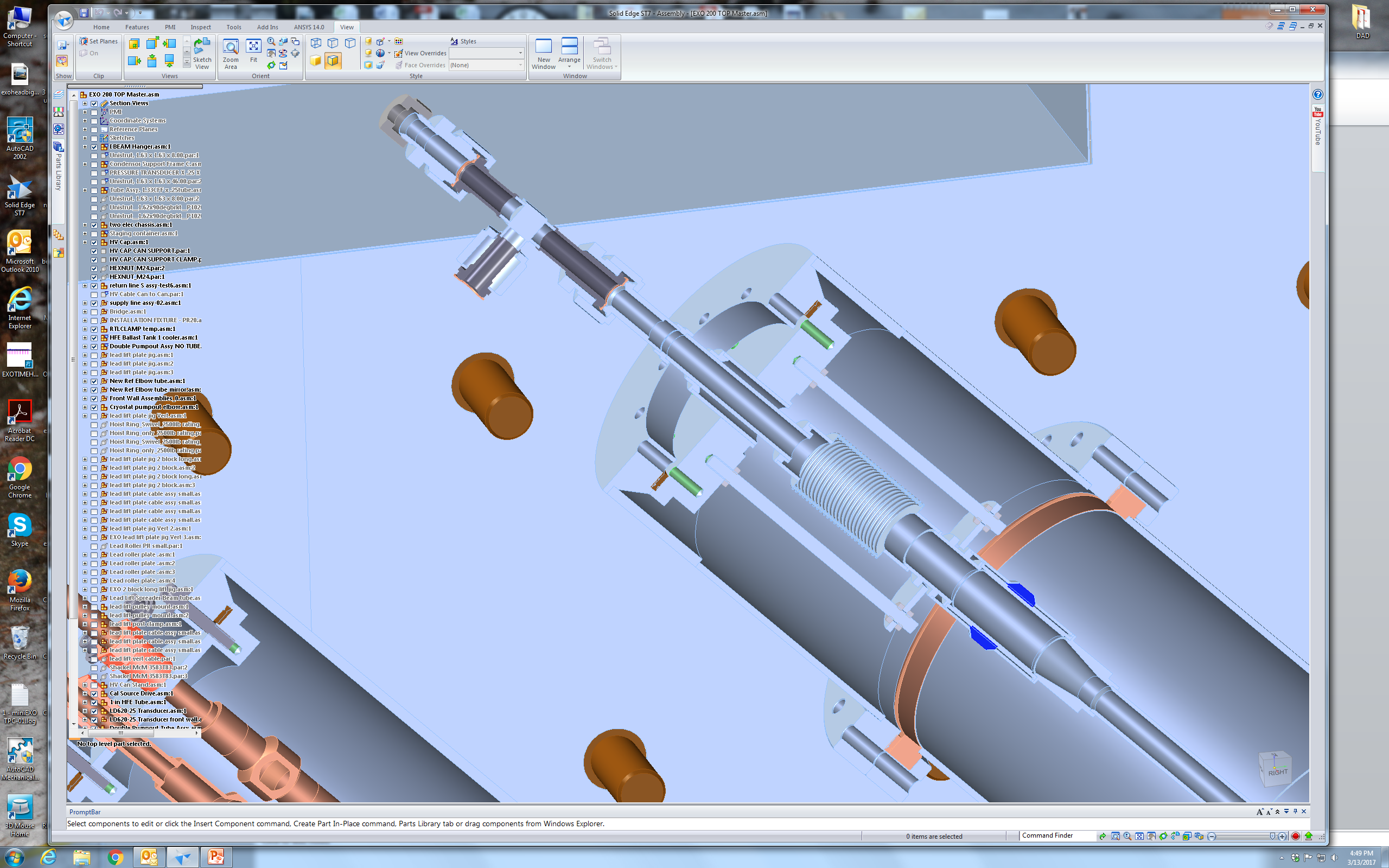}
\caption{Left: Front view of the installed cryostat with no lead shielding in place.  The cathode high-voltage connection (upper left), refrigerant connections (left), cryostat thermocouples (lower left), E-boxes for front end electronics (bottom), and the supply transfer line (right) are shown.  Right: CAD detail of the feedthrough for the cathode HV cable.  The air side of the cable is in the upper left and the connection to the xenon space is in the lower right.  During assembly the outer vacuum jacket fits over the airside connector, requiring the central through hole and reverse CF seal. }
\label{fi:cs}
\end{center}
\end{figure}

All services incorporated right-angle bends outside the inner lead shield to keep non-copper components out of line of sight of the TPC \cite{Auger:2012gs}. The refrigeration, high voltage, and HFE lines continued the insulating vacuum through to feedthroughs after brazed transitions to stainless steel for the vacuum jacket.  
These feedthroughs were mounted on reverse CF flanges (see figure \ref{fi:cs}), in which the bolt circle is closer to the center than the sealing gasket, with a central through hole to allow these large rigid assemblies to be sealed from the outside.

The OV was designed to withstand vacuum inside without significant deformations. This required mounting a reinforcement disk to the back plate by electron beam welding from the outside. The weld was ``by transparency'', fully penetrating the plate.  The IV was designed to withstand vacuum inside, as well as an overpressure of 200~kPa to support the explosive load from the TPC (see section \ref{pcon}).  The deformations under all these conditions were verified in pressure tests with a demineralized water filling.

As discussed in \cite{Auger:2012gs}, the IV was wrapped in embossed superinsulation \cite{sheldahl} measured to have acceptable levels of radioactive contamination.  The choice of embossed material avoided the gauze between layers that was found to be too radioactive.   Five layers of \SI{6.35}{\um} thick superinsulation were used, a reduced total quantity chosen to minimize radioactivity contributions while retaining sufficient performance at \SI{170}{\K}.   The total heat leak into the IV, arising from conduction through the refrigeration line feedthroughs, the HFE fill lines, the teflon support blocks, the vacuum space, the xenon transfer lines, and the signal cables, was found to be about \SI{90}{\W} from the rate of temperature rise without cooling.  The radiation contribution to this heat load was shown to be negligible based on its temperature dependence determined from rate-of-rise data collected at two cryostat temperatures.

\begin{figure}[ht]
\begin{center}
\includegraphics[height=6cm]{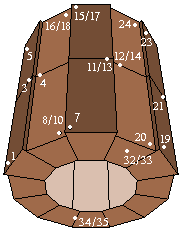}
\includegraphics[clip=true,trim=1cm 1cm 1cm 0cm,height=6cm]{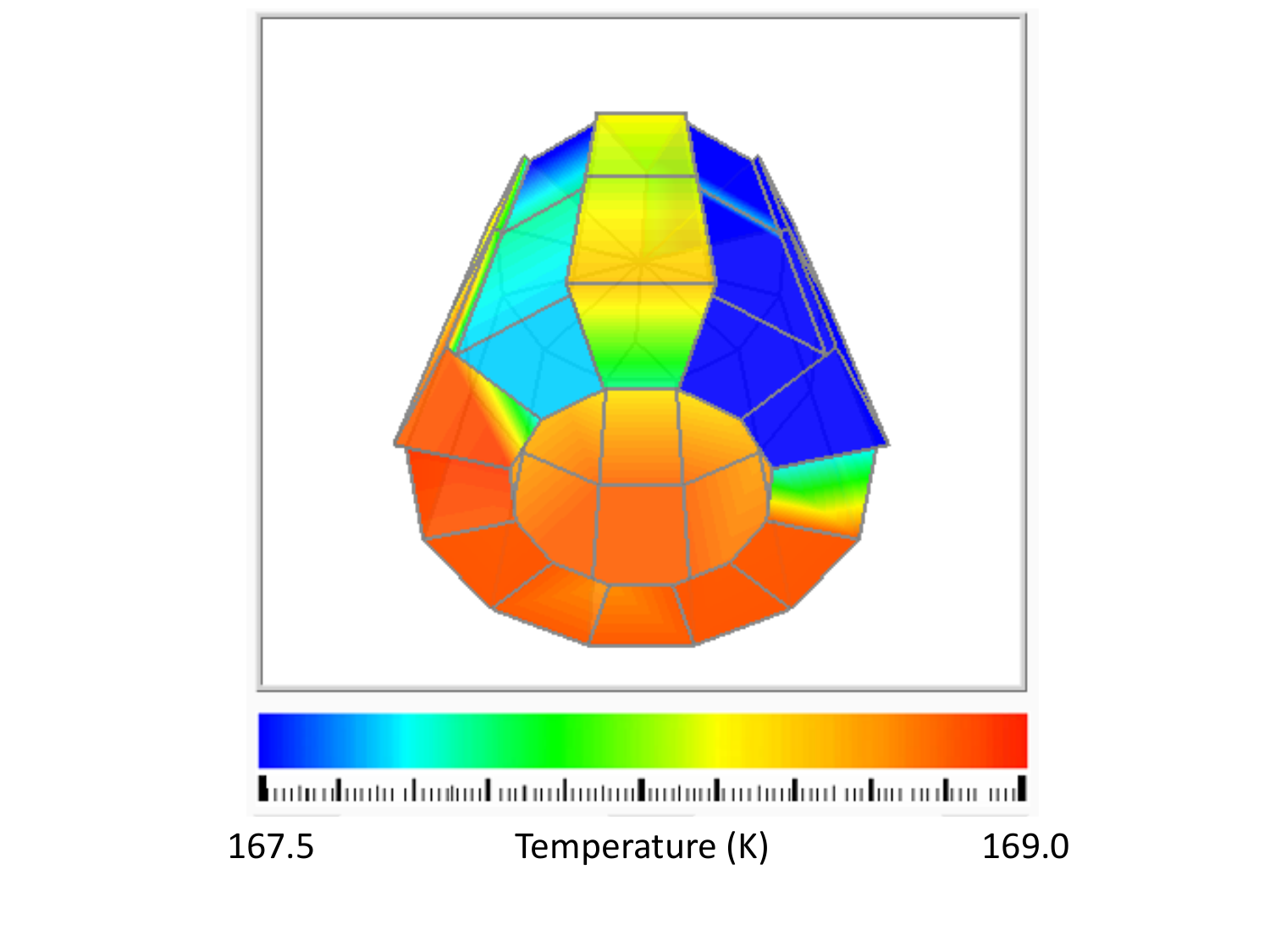}
\caption{Left: The location of the thermocouples on the upper side and in the front of the IV; right: temperature distribution during routine operation with the heat exchanger at 2 o'clock in use.}
\label{fi:IV_T-dist}
\end{center}
\end{figure}

\subsection{Cooling and Temperature Control}
\label{sec:temp_control}
In the prototype, HFE cooling was accomplished via an immersed refrigeration coil, which had the drawback of cooling power loss as frozen HFE accumulated on the coil.  In EXO-200, the refrigeration was accomplished with three heat exchangers (HXs) welded on the outer surface of the IV, as shown in Figure \ref{fi:IV_HX}.  This arrangement avoided the HFE ``icing" issue because of the direct thermal contact with the large copper mass.  Each HX was made out of a portion of a NOSV copper plate. Two grooves with (1$\times$\SI{1}{\cm^2}) cross-sections were machined on the inner side. They served as channels for refrigerant and were connected to circular holes at the front end of the HXs. The original design allowed for two redundant loops per HX, but leaks between loops in each pair resulted in the decision to jumper the loops together.  Each cryostat refrigerator (section \ref{sec:fridges}) supplied refrigerant to a single HX via vacuum-jacketed transfer hoses \cite{vbs}.

The HXs were transparency-electron-beam welded to the corresponding side plates of the IV before assembly of the vessel.  Pressure tests with nitrogen after assembly showed measurable deformations at \SI{2}{\MPa}, while a \SI{3}{\MPa} rating was deemed necessary to withstand possible pressure surges. The HXs were therefore reinforced with additional MIG welds to the IV in three grooves machined between and parallel to the fluid circulation channels.  Each groove was \SI{10}{\mm} wide at the bottom and \SI{40}{\mm} wide at the top prior to welding, and the copper used in the MIG welds was tested for low radioactivity.  The HXs withstood the full pressure test after reinforcement.

\begin{figure}[ht]
\begin{center}
\includegraphics[clip=true,trim=0cm 0cm 0cm 5.6cm,height=5cm]{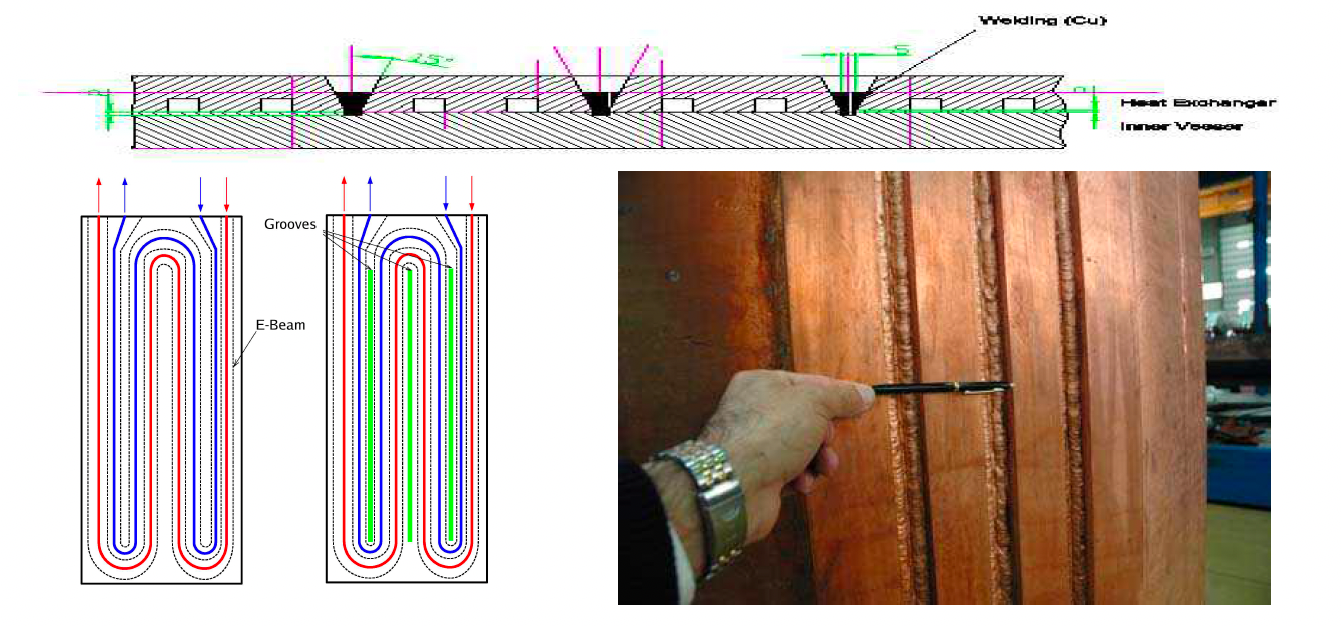}
\caption{Scheme of the reinforcement of the HXs (left) and views of the results (right). }
\label{fi:IV_T-dist}
\end{center}
\end{figure}

All three HXs were initially connected to refrigerators, but leak-free connections to the soft pure copper cryostat proved difficult to achieve.  Some locations slowly leaked refrigerant, which was detected by a residual gas analyzer (RGA) monitoring the cryostat vacuum space.  Eventually, one HX (at the 10 o'clock position) was retired from service and valved closed.  The copper refrigerant lines in the vacuum space were brought out through two feedthroughs, one of which carried two refrigeration circuits.  In routine operation one HX was sufficient to maintain the cryostat temperature within specifications.

The IV was instrumented with 40 type T thermocouples (TCs) to monitor the temperature distribution. They were made from constantan and copper wires ($\oslash =\SI{80}{\um}$) plasma-welded together, electrically insulated with a thin sheet of teflon, and screwed against the IV exterior. The location of the thermocouples and a typical temperature distribution are shown in Figure \ref{fi:IV_T-dist}, with only one HX, at the 2 o'clock position, operational. It is seen that the temperature was spatially uniform around \SI{168.2}{\kelvin} to within \SI{0.7}{\kelvin}.  The coldest temperature seen on any TC during operation was greater than \SI{160}{\kelvin}.

The prototype system at SLAC employed "ON/OFF" temperature control via the refrigerator cooling valve (see section \ref{sec:fridges}), resulting in measured peak-to-peak temperature swings of \SI{1}{\kelvin}.  Control using additional cryogenic proportional valves was attempted both in the prototype and later as a retrofit to EXO-200 but failed to show linear behavior.  This might indicate that the refrigerant flow rate was dominated by large impedances internal to the refrigerators rather than by the setting of the external proportional valve.  The cryostat FEA simulation described below indicated the "ON/OFF" temperature swings in the full cryostat would be much smaller than in the prototype, so that scheme was adopted for EXO-200 as well.   

Cryostat cooling was regulated though the slow control (section \ref{sec:SC}).  Three of the thermocouples instrumenting the IV were designated as the control thermocouples for this process.  One of these three was designated as the primary and normally used in the cooling algorithm, however a voting process using all three control thermocouples could change that selection.  For the majority of detector operation, the primary TC was one located on the center of the IV's back plate.  This choice of a control sensor on the cryostat's central axis minimized shifts in the average cryostat temperature when different refrigerators, coupled to different heat HXs, cooled the cryostat.  A \SI{0.05}{\kelvin} hysteresis band was typically included in the "ON/OFF" cooling algorithm applied to these thermocouples, resulting in \SI{0.2}{\kelvin} peak-to-peak variation in their readings.  The actual temperature variation within the IV was measured to be \SI{0.04}{\kelvin} peak-to-peak during commissioning using temperature sensors immersed directly in the HFE and later removed for data collection because of radioactivity concerns.

\subsection{Radioactivity Control}
When designing and building the cryostat, great attention was paid to minimizing the background from internal activities.  All components, namely the NOSV copper, the seals, the copper for the MIG welding, the PTFE for the blocks, the superinsulation material, the thermocouples, all fasteners, and silicone lubricant used on fasteners, were chosen for their low intrinsic radioactivity \cite{Background_NIM}. After purchase, the plates were stored underground in the Gotthard and the Vue-des-Alpes underground laboratories until construction began in order to reduce cosmogenic activation. On site at SDMS, to further reduce cosmogenic activation, the plates and the subcomponents already made were stored in a well with a \SI{1}{\m} concrete overburden, providing some protection against neutrons. 

To avoid contamination by contact, workers were required to wear gloves when handling components. When clamping was necessary, the parts in contact were protected with thin sheets of clean copper. Whenever possible, the parts were protected with a sheet of plastic. Cutting was done with water jets at high pressure.  An additional \SI{15}{mm} of material was removed afterward by machining to avoid any contamination from the abrasive powder included in the cutting jets.  Machining was performed with clean tools using methanol for lubrication. Welds were made by electron beam without the addition of foreign material, except for the reinforcements to the heat exchangers described in Section \ref{sec:temp_control}.

The completed vessels were thoroughly cleaned, including acid etching following removal of loose debris, after receipt from SDMS.  A solution of 30\% HNO$_3$ was sprayed on all surfaces, and after 15 minutes the vessels were rinsed with demineralized water. This operation was repeated twice.

\subsection{Thermal Simulations}

\begin{figure}
\begin{center}
\includegraphics[height=6.0cm]{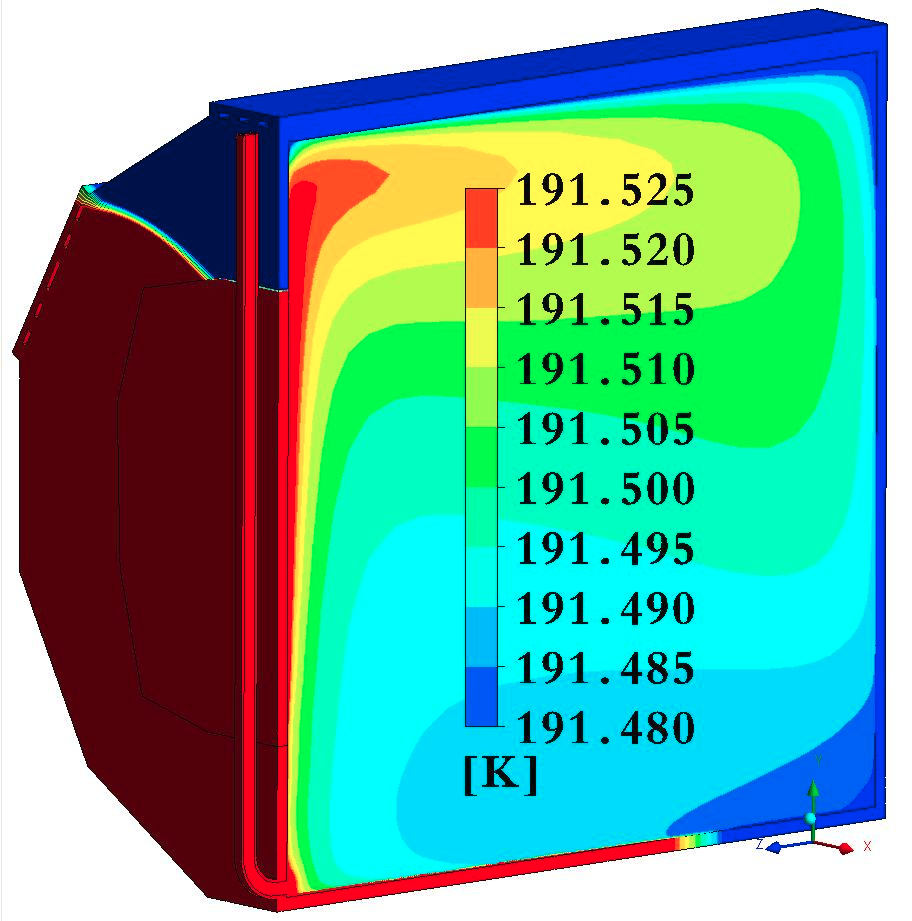}
\includegraphics[height=6.0cm]{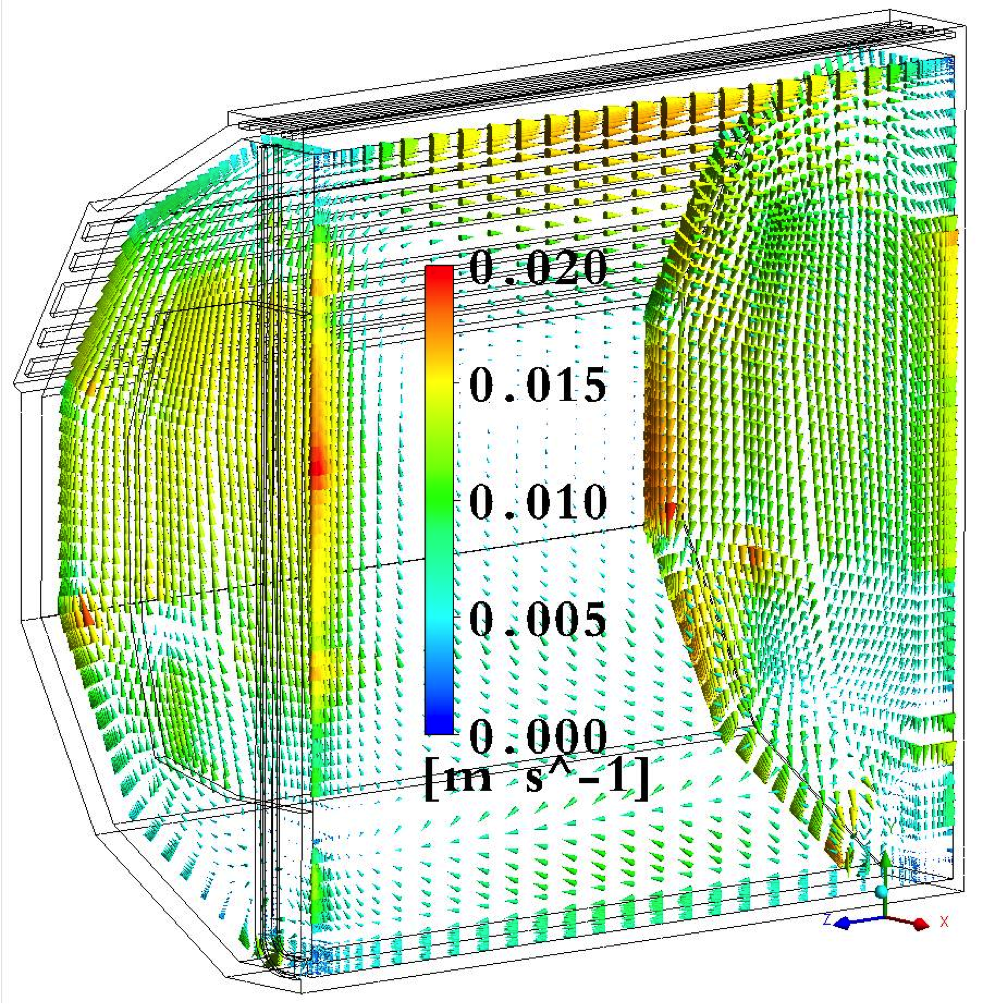}
\end{center}
\caption{Temperature profile (left) and velocity vectors showing the flow (right) in the HFE at one point in the cooldown simulation.  The highest thermal gradients are in the boundary layers near the copper surfaces, while those in the bulk HFE are minimal.}
\label{fig:sim3}
\end{figure}

The success of the cryostat scheme depended on efficient heat transfer through the HFE, primarily through natural convection, to the cryostat and its attached HXs, both to limit spatial and temporal temperature gradients at base temperature and to provide for a reasonable cooldown time from room temperature.  Numerical simulations were used initially to show that this concept was viable and to guide design choices.  Later, detailed simulations of the final design implementing measured heat leaks and known refrigerator cooling power reproduced the thermal behavior of the cryostat during initial commissioning.  All models were half-symmetric so the HXs on the side flats of the cryostat were forced to have equal cooling power in the simulation.  An example from a cooldown simulation is shown in Figure \ref{fig:sim3}.

The Computational Fluid Dynamics (CFD) program CFX, which is part of the ANSYS suite of multiphysics analysis programs \cite{ANSYS}, was used to simulate heat transfer in conjunction with fluid flow.  Called conjugate heat transfer, this form of CFD allows one to include solid regions with no flow within the CFD model.  CFX uses a finite volume method to solve the Navier-Stokes equations for fluid flow.  Due to convergence issues, the meshes created for the finite volume analyses were all linear extrusions of two-dimensional meshes.  All elements were hexahedrons.  Anomalous flow effects were reduced by limiting the number of points in the free mesh where five elements share a corner node and positioning those points where the flow in the plane of the free mesh is minimal.  Material properties for HFE were supplied by the manufacturer.  Analysis of the cryostat cooldown required a full buoyancy treatment of the problem, whereas the Boussinesq approximation (see \cite{boussinesq} for a contemporary review) was used for analysis of the steady-state operation with constant HFE density. 

\subsection{Refrigerators}
\label{sec:fridges}

The Polycold 672HC refrigerators originally used to cool the EXO-200 cryostat employed a five-component hydro-fluorocarbon/argon gas refrigerant mixture and a cascaded refrigeration cycle, producing a minimum temperature around \SI{120}{\kelvin} and output ("feed") temperatures typically just above \SI{130}{\kelvin}. The cold refrigerant delivered to the external circulation loop consisted almost entirely of tetrachloromethane (R-14) and argon, with the less volatile components remaining condensed within the unit.  Each HX in use contained as much as \SI{1}{kg} of this mixture, the radiopurity of which was not measured.  These high-capacity refrigerators were rated for \SI{1500}{\watt} cooling power at room temperature and included a "buffer/unloader" feature that periodically diverted refrigerant flow during periods of high load in order to prevent extended overpressure conditions at the discharge side of the compressor.  Smooth cooldown from room temperature required three days with two refrigerators operating in parallel.  A single unit maintained the cryostat temperature at its setpoint around \SI{168}{\kelvin} against the relatively small \SI{90}{\watt} heat leak. 

The Polycold refrigerators operated for about seven years of commissioning and data-taking.  While they were generally reliable, regular checks of the refrigerant pressure at room temperature showed a slow but steady loss, so that "top-off" refrigerant charges were added at approximately two-year intervals.  This loss was probably caused by small refrigerant leaks at the connections to the cryostat HXs.  Occasionally more dramatic leaks were observed and addressed in the units themselves or at the connection to the refrigeration lines.  

The combination of such events with the general age of the Polycold units resulted in their replacement with Telemark \cite{telemark} TVP 1800 refrigerators, which were nominally very similar in operational principles and capabilities.  The Telemarks were successfully used for the remainder of detector operations, though they provided about 25\% less cooling power than the Polycolds as judged by cooldown rates and duty cycle at setpoint, despite nominally being 20\% more powerful.  In addition, the internal compressors in the Telemarks, unlike those in the Polycolds, did not incorporate check valves, so that in one case refrigerant was cryopumped from a Telemark to the cold cryostat HX when it was warmed up for an internal pressure check.  Check valves were added to the external refrigerator plumbing to prevent that from recurring.

Cryostat cooling cycles were observed to cause excess noise events in the collected data, typically during the start of a cooling cycle.  This noise was in the form of $\sim$kHz signals on particular data channels, and was mainly caused by microphonics originating in the refrigerant lines connecting to the cryostat feedthroughs near the front-end electronics.  While the Polycold refrigerators were in use, the rate of such noise events varied significantly in time, often unobserved completely and other times requiring several minutes of data to be discarded for each cooling cycle.  With the Telemark refrigerators in use, the rate of these events was larger and consistent in time, and at least \SI{1}{\minute} of data was cut for each cooling cycle.  The typical time between the start of cooling pulses was \SI{75}{\minute} with the Telemark refrigerators cooling the cryostat.  The reason for the difference in the DAQ noise produced by the two types of refrigerators is not known.

 \subsection{Chilled water system}
 \label{sec:TS}
 
Chilled water was not available in the WIPP underground, so air-cooled water chillers were used to supply the refrigerators.  Initially BV Thermal Systems \cite{bvthermal} units were used, later replaced by Coolant Chillers \cite{coolant_chillers} units.  Originally, each refrigerator was supplied by a dedicated water chiller, and one refrigerator and one chiller were supplied with UPS backup power (see section \ref{sec:EP}) to maintain cryostat cooling in the event of a power outage.  This provision was found to be inadequate, however, in the event of a site-wide power outage that would also bring down the ventilation underground, since the lack of ventilation would render even a UPS-powered water chiller ineffective.  There was also no provision for cooling the UPS units during a power outage, reducing confidence in their performance.  Finally, the original configuration was also difficult to adjust to changing conditions, for example when the UPS-powered fridge or chiller needed to be turned off for maintenance, or if a chiller suddenly failed.   

\begin{figure}
	\begin{center}
        \includegraphics[width=5in]{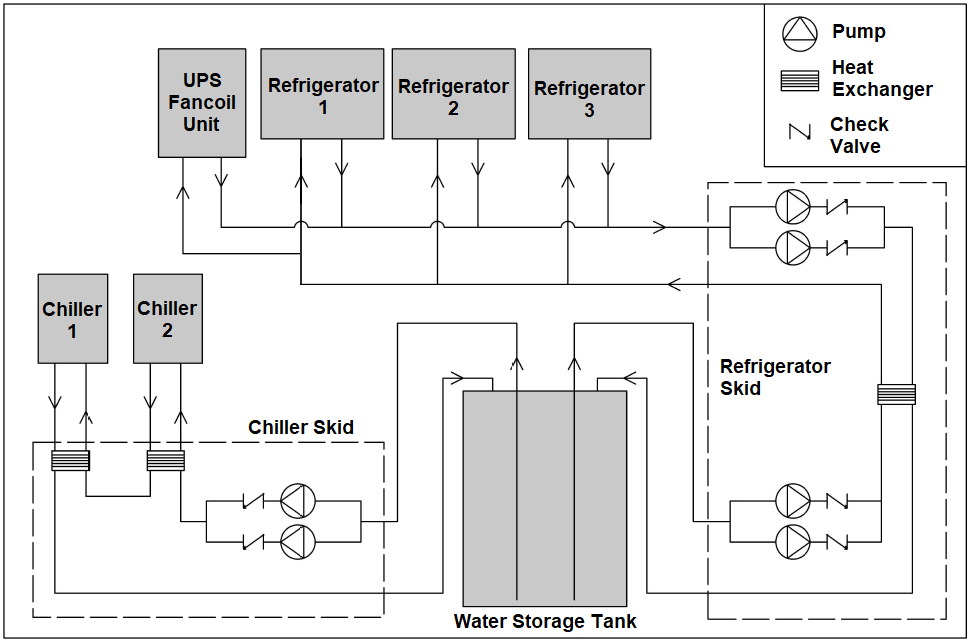}
	\caption{Simplified schematic of the final chilled water system, in which the heat capacity of the stored water absorbed waste heat during power outages.  \label{fig:thermalstore}}
         \end{center}
\end{figure}

To address these concerns, a "thermal store" was incorporated into the chilled water system (see Figure \ref{fig:thermalstore}).  In the final system, the thermal store was a \SI{11}{\meter^3} double-walled plastic water tank.  In normal operation, water circulated by skid-mounted pumps located next to the tank carried heat from the refrigerators to the thermal store water through an intermediate heat exchanger, while a pair of water chillers cooled the tank water through another heat exchanger.  When power was lost, the water chillers and chiller-skid water pumps shut down, while the UPS-powered refrigerator and refrigerator-skid water pumps continued to operate.  In this way, the thermal store water continued to absorb the refrigerator waste heat during power outages.  A fan coil unit was also installed in the UPS container and supplied with thermal store water to maintain temperature control during a power outage.  The thermal store system was instrumented for constant monitoring via the slow control system.  Thermistors monitored the tank water temperature at various depths as well as both heat exchangers, and the water flow in each loop was measured with a flow meter.

 The cryostat refrigerators could tolerate at least a\SI{12}{ \degreeCelsius} increase in chiller water temperature over the normal operating temperature, so that the total useful heat capacity in the thermal store was about \SI{0.57}{\giga{J}}.  While the system was not tested to failure, power outages as long as \SI{12}{\hour} occurred at least annually.  The observed rate of warming during these outages, about \SI[per-mode=symbol]{0.5}[]{\degreeCelsius\per\hour}, was consistent with the system's ability to support \SI{24}{\hour} of cryostat cooling without electrical power.

\section{Xenon System} 
\label{sec:xesystem}

\begin{figure}
	\begin{center}
        \includegraphics[width=5in]{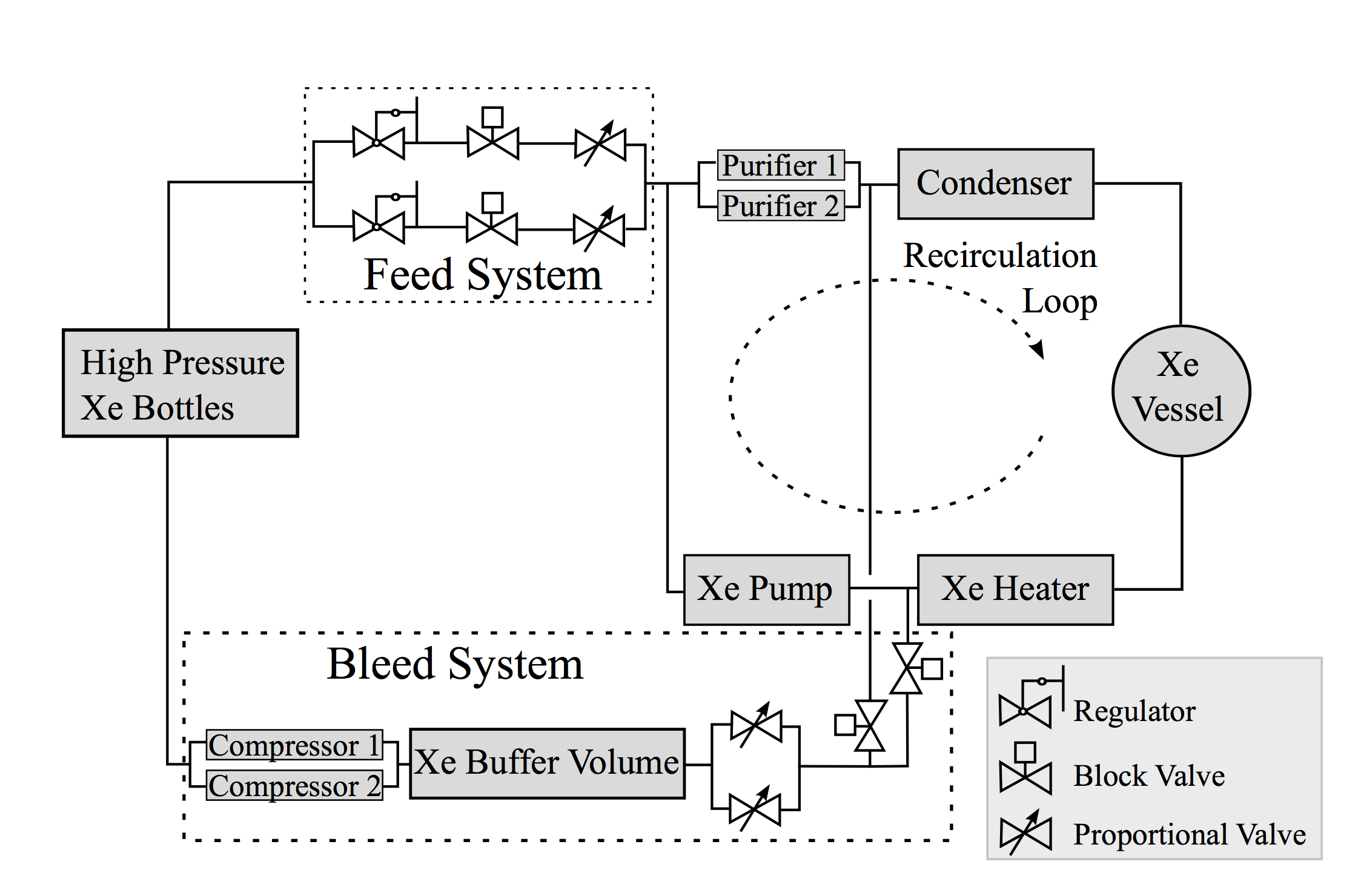}
	\caption{A high-level diagram of the xenon system, including the the recirculation loop for xenon purification and the feed/bleed system to limit the pressure differential across the TPC vessel.  \label{fig:xedgm}}
         \end{center}
\end{figure}

The primary functions of the xenon system were transfer of xenon between the gas bottles and the LXe vessel, purification of the LXe, and continuous regulation of the thin-walled TPC vessel dP.  The key elements of the system, shown schematically in Figure \ref{fig:xedgm}, were the feed and bleed systems that moved Xe gas in and out of the system, respectively, and the recirculation loop that vaporized, purified, and recondensed LXe. 

\subsection{Feed and Bleed}
\label{sec:xetransfer}

\begin{figure}
	\begin{center}
        \includegraphics[width=5in]{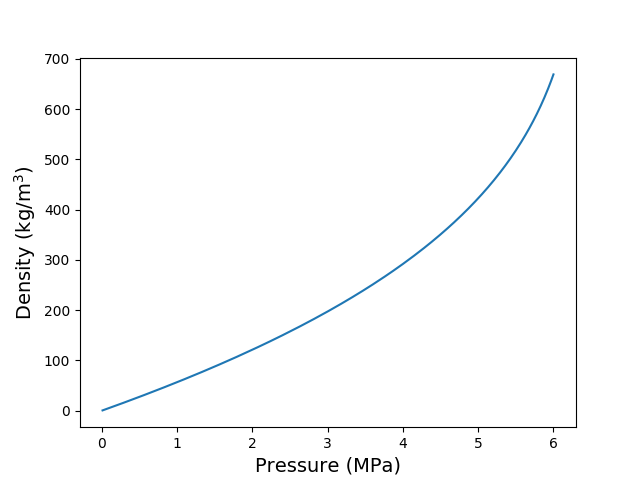}
	\caption{Xe density vs pressure at \SI{295}{\kelvin}.  Data were taken from \cite{nist_thermophysical}. \label{fig:xe_density}}
        \end{center}
\end{figure}

The feed system moved xenon to the TPC vessel from a rack of storage bottles connected in parallel and kept open throughout detector operation.  The bottle rack pressure ranged from about \SI{5.6}{\MPa} when the entire stockpile was in storage down to about \SI{0.5}{\MPa} when the vessel was filled.  A plot of Xe density vs pressure is shown in Figure \ref{fig:xe_density}.  The bottle rack supplied two redundant feed circuits.  In each circuit a regulator \cite{regulators} reduced the xenon pressure to around 0.21 MPa, and a computer controlled block valve and proportional valve in series regulated the flow of xenon into the recirculation path. The pressure rating of the block valves exceeded \SI{6.9}{\MPa}, allowing them to hold off bottle pressure in case of catastrophic regulator failure.

The bleed system moved xenon from the vessel back to the gas bottles. The central component of the bleed system was a redundant pair of Fluitron two-stage compressors \cite{compressors}, each with a compressing rate of \SI[per-mode=symbol]{53}{\liter\per\minute} for input pressures greater than \SI{200}{\kPa}.  The flow rate dropped to \SI[per-mode=symbol]{10}{\liter\per\minute} at \SI{100}{\kPa} and to zero at \SI{80}{\kPa}.  The output of the compressors was connected directly to the bottle rack.  The input of the compressors was connected to a \SI{50}{\liter} buffer volume, and they operated as necessary to maintain the pressure of that buffer volume within programmed limits.  In addition, a hardware switch could activate the compressors directly if the pressure exceeded \SI{87}{\kPa}.   The buffer volume was connected to LXe system by a pair of parallel proportional valves followed by a pair of parallel block valves.  These parallel bleed circuits tapped the circulation path between the xenon heater and the xenon pump and between the purifiers and the xenon condenser, respectively.   Bleed capability was provided for situations where the system pressure was below \SI{80}{\kPa} (for example when emptying the detector) by a cryopump bottle immersed in LN$_2$ and connected to the buffer volume.  Swagelok stainless-steel sample cylinders \cite{cryopump} with VCR to NPT adapters welded in place were used for this purpose.

\begin{figure}
\begin{center}
\includegraphics[width=0.9\textwidth]{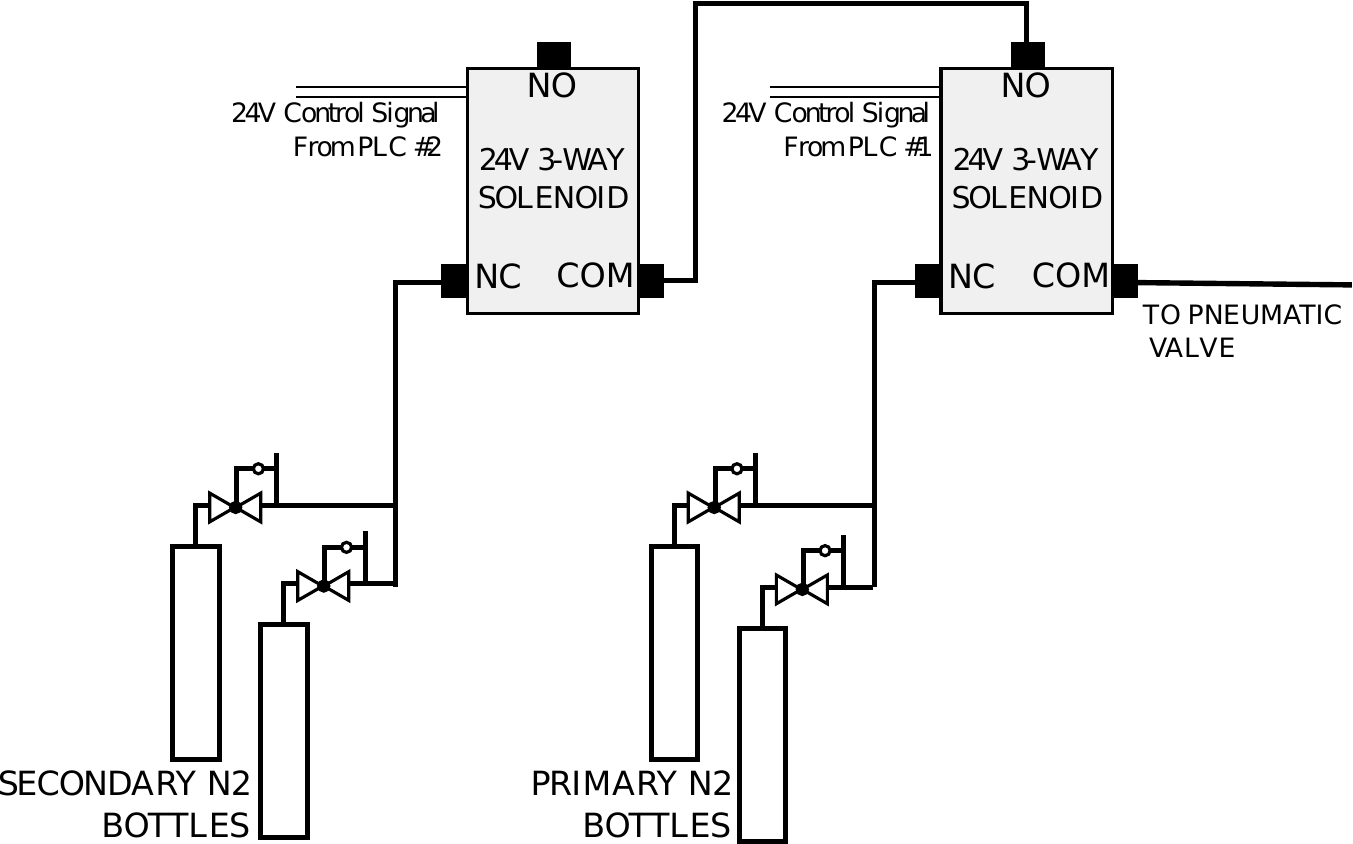}
\end{center}
\caption{Connection scheme for pneumatic system solenoid valves.}
\label{fig:solenoid}
\end{figure}

The feed and bleed pneumatic block valves were actuated by pairs of small three-way solenoid valves connected in the configuration shown in Figure \ref{fig:solenoid}.  Each of the two solenoid valves received compressed nitrogen from a set of two bottles with regulators attached.  One regulator was set slightly higher than the other so that one bottle would exclusively provide the gas until it emptied, at which point the other bottle, still full, would provide gas until the first bottle was replaced.  Finally, each of the two solenoid valves was controlled by a different slow control PLC (see section \ref{sec:SC}).   This arrangement was intended to maintain pressure control despite the failure of a solenoid valve, the emptying of a nitrogen supply cylinder, or the failure of one of the slow control PLCs.  The HFE system (section \ref{sec:HFE}) also included three pneumatic valves used mainly for cryostat overpressure protection that were controlled by solenoid valves in this same arrangement. 

\subsection{Pressure Control}
\label{pcon}

The feed and bleed systems were operated through the slow control (section \ref{sec:SC}) and acted automatically to keep the TPC vessel dP within a programmable deadband of the chosen target value.   Thus the Xe system pressure was adjusted relative to the HFE system pressure, rather than to an independent setpoint.  The settings of the proportional valves during feed or bleed events were determined by PID algorithms incorporated into the slow control software.  In practice, however, the proportional valves opened fully during most feeds and bleeds, so that the system effectively operated in an on/off mode.  Because of the importance of limiting dP, substantial effort was made to keep this function robust against component failure, including the use of redundant feed and bleed circuits and the use of multiple pressure gauges.

\begin{figure}
\begin{center}
\includegraphics[clip=true,trim=2cm 0cm 0cm 0cm,height=9cm]{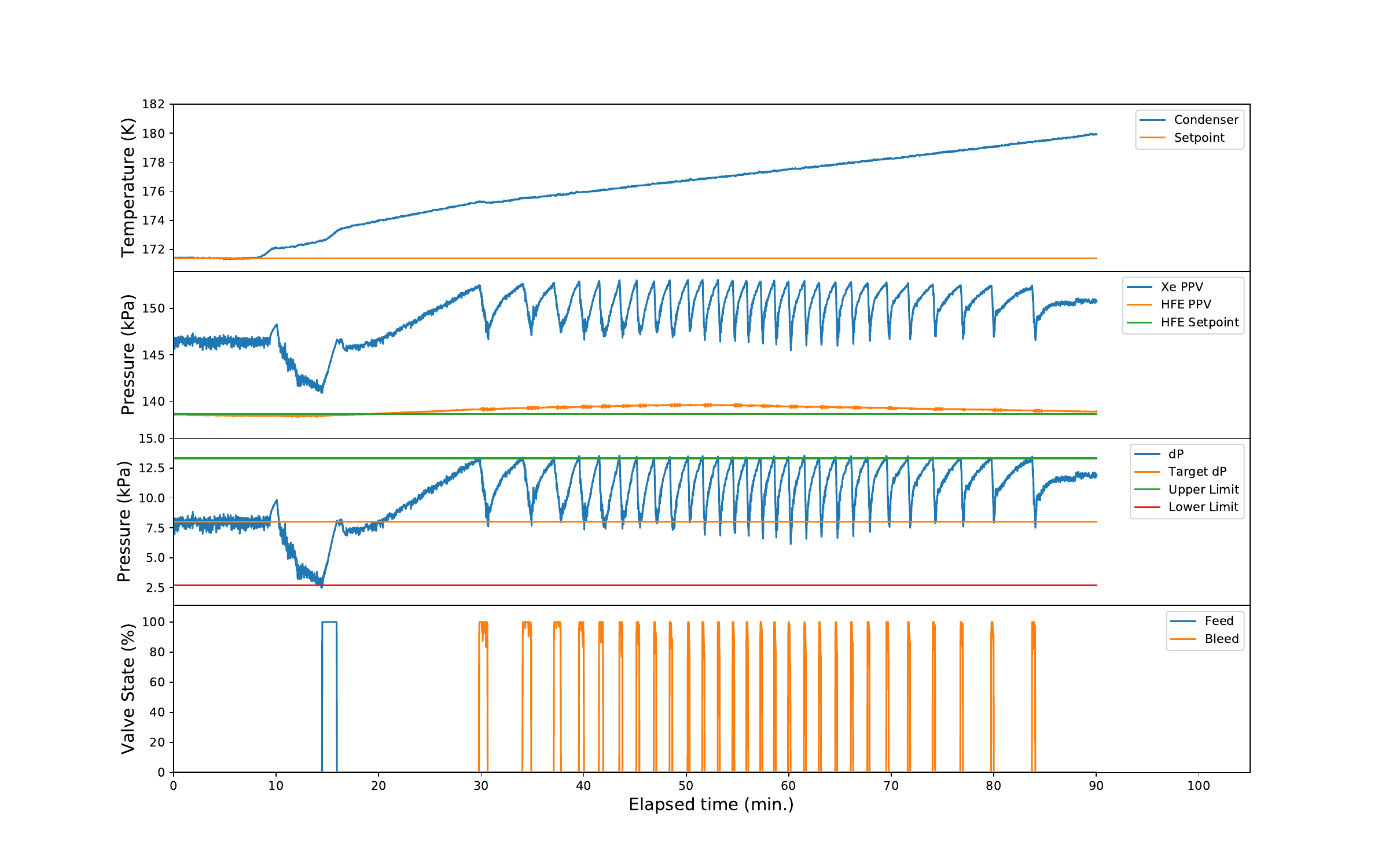}
\caption{An example of pressure control during an unplanned power outage.  Pressure regulation by the Xe heater is interrupted when power is lost at around \SI{8}{\minute}.  The recirculation interlock (see section \ref{sec:recirc}) turns off the recirculation pump and Xe heater while the Xe condenser slowly warms (top plot).  Initially the Xe system PPV (second plot from top), and therefore the TPC vessel dP (third plot from top), falls without evaporated gas to offset condensation, and the system feeds Xe gas (bottom plot) to keep dP above the lower limit.  As the condenser continues to warm, condensation stops and the \SI{0.4}{kg} of LXe in the supply line vaporizes, raising the Xe PPV and therefore dP.  The system then bleeds Xe gas to keep dP below the upper limit.  Once the supply line is empty, around the \SI{90}{\minute} mark, the system pressures stabilize.}
\label{fig:feedbleed}
\end{center}
\end{figure}

The target dP and the size of the feed/bleed deadband were adjusted as needed during various detector operations.  The most extensive use of the feed and bleed systems came when filling the detector with or emptying it of liquid Xe (see section \ref{transfer}), since those operations transferred the majority of the Xe stockpile, and during those operations the target dP was typically 0 with a deadband around $\pm$\SI{2}[]{\kPa}.  During normal operation with the detector full and recirculation (see section \ref{sec:recirc}) underway, the target dP was \SI{8.1}{\kPa} with a $\pm$\SI{5.3}[]{\kPa} deadband.  Since Xe feeds negatively affected the Xe purity (section \ref{purity}) and Rn levels (section \ref{sec:radon}), the wider feed/bleed deadband allowed the Xe heater (section \ref{sec:recirc}) to provide fine pressure control without impacting data collection, while the feed and bleed systems could still activate to protect the TPC in upset conditions like power outages (see Figure \ref{fig:feedbleed} for an example).  The typical variation in dP under this fine pressure control was $\pm$\SI{0.67}[]{kPa}.

For the purposes of pressure control, dP was defined as the difference between the Xe pressure at the TPC vessel midplane and the HFE pressure at the same point.  These two pressures were designated the Xe and HFE system pressure process variables (PPVs) respectively.  The Xe PPV was computed from direct pressure measurements in one of the transfer lines (see section \ref{sec:recirc}) either above or below the vertical level of the TPC vessel, with a correction applied for the liquid head pressure determined from the difference in those measurements.  The pressure transducers used for these pressure measurements were MKS Baratron \cite{baratrons} capacitance manometers.  The Xe PPV was typically \SI{147}{kPa} during normal operation, including a \SI{16}{kPa} liquid head and \SI{131}{kPa} gas pressure.  This value of the system pressure was found to be conveniently above the  \SI{121}{kPa} vapor pressure of Xe at the \SI{168.2}{K} cryostat temperature (see Figure \ref{fig:xe_phase} for a Xe phase diagram).  The HFE PPV is discussed in section \ref{sec:HFE}.  An additional redundant dP measurement was derived from a differential pressure sensor \cite{rosemounts} connected to both systems with a correction applied for the difference in the Xe and HFE liquid heads.  The immediate connection of that sensor to each system was made by a stainless-steel diaphragm \cite{picseal} connected in turn to the differential pressure sensor via an intermediate fluid-filled line.

A recovery balloon constituted a last resort to limit the Xe system pressure and to prevent the loss of the Xe stockpile in scenarios where pressure control was lost.  The balloon was made from \SI{51}{\micro\meter} flame-retardant polyurethane and was sized to contain \SI{200}{kg} of $^{136}$Xe at the temperature and pressure of the WIPP underground (see section \ref{sec:EI}).  The balloon was hung in a nearby drift alcove and plumbed to the Xe system inside the cleanroom through CPVC pipe.  The balloon plumbing connected to the Xe system behind a block valve in series with a relief valve.  The block valve could be actuated through the slow control interface, but no logic existed to do so automatically.  The relief valve, which would open at a pressure \SI{83}{kPa} above atmospheric pressure, ensured that opening the block valve would only transfer Xe to the balloon if the system pressure was very significantly elevated.  The block valve was configured to be normally-open, so that this scheme would still be effective in a scenario where the slow control lost electrical power.  The balloon was also connected to five burst disks providing overpressure protection at various locations in the system.  No Xe was ever transferred to the balloon.

The TPC vessel dP was kept within its $\pm$\SI{35}[]{kPa} design limit through a wide range of normal and off-normal conditions over the course of the experiment with two exceptions.   In the first case, the unforseen combination of sub-system failures resulting from the failure of an auxiliary power UPS unit during and after a power outage produced a maximum dP of \SI{142}{kPa} (see section \ref{sec:EP}).  In the second case, a \SI{-80}{kPa} dP occurred following the 2014 WIPP incidents (section \ref{events}) amid a power outage in conditions including a leak in the Xe system, loss of HFE system pressure control (section \ref{sec:HFE}), and no underground access.  The LXe had been recovered to storage about seven months before the latter event, soon after the WIPP incidents themselves.  Regular source calibrations used to monitor detector performance \cite{prc} indicated that neither event damaged the TPC.

\subsection{Recirculation}
\label{sec:recirc}

In order to maintain the Xe purity (section \ref{purity}) critical to TPC operation, the Xe was constantly recirculated through a pair of commercial gas purifiers connected in parallel.  This required an external heater and condenser to process the LXe.  These elements, together with a gas pump to overcome the flow impedance of the purifiers, constituted the recirculation loop.  The feed system tapped into this loop immediately before the purifiers, so that feed gas passed through them as it entered the system. This placement was also advantageous in cleansing impurities during filling of the vessel.  

The recirculation originally was envisioned to be carried out through a single purifier with the parallel unit as a backup, and the specifications of the SAES \cite{SAES} model selected for use determined a maximum design recirculation rate of \SI{20}{SLPM}.   The flow impedance of the purifiers was found to be larger than expected, however, and so the recirculation was instead carried out through both units to avoid exceeding the maximum differential pressure that the recirculation pump could provide.  Despite the use of two purifiers instead of one, the maximum recirculation rate was still limited to \SI{20}{SLPM} by the capabilities of the other recirculation components.

The condenser was composed of \SI{6.6}{m} of 3/8 inch copper tubing wound around the upper half of a copper block and brazed to it. A second coil installed around the lower half of the copper block carried refrigerant supplied by a dedicated refrigerator.  A heater placed in between the coils was used to regulate the temperature of the upper half of the block. Ten type-T thermocouples measured the temperature in various places on the copper block and coils. One of these, selected from among three by the slow controls software in a voting process, was used in a PID-algorithm feeding back to the trim heater.  The use of three control TCs was intended to protect against the malfunction of any one, though in practice it was additionally found necessary to require reasonable values from each control TC before allowing it into the voting process.  It was also found necessary to account for temperature offsets between the TCs in order to maintain smooth control when a new one was selected for the PID feedback.

\begin{figure}
	\centering
	\includegraphics[width=5in]{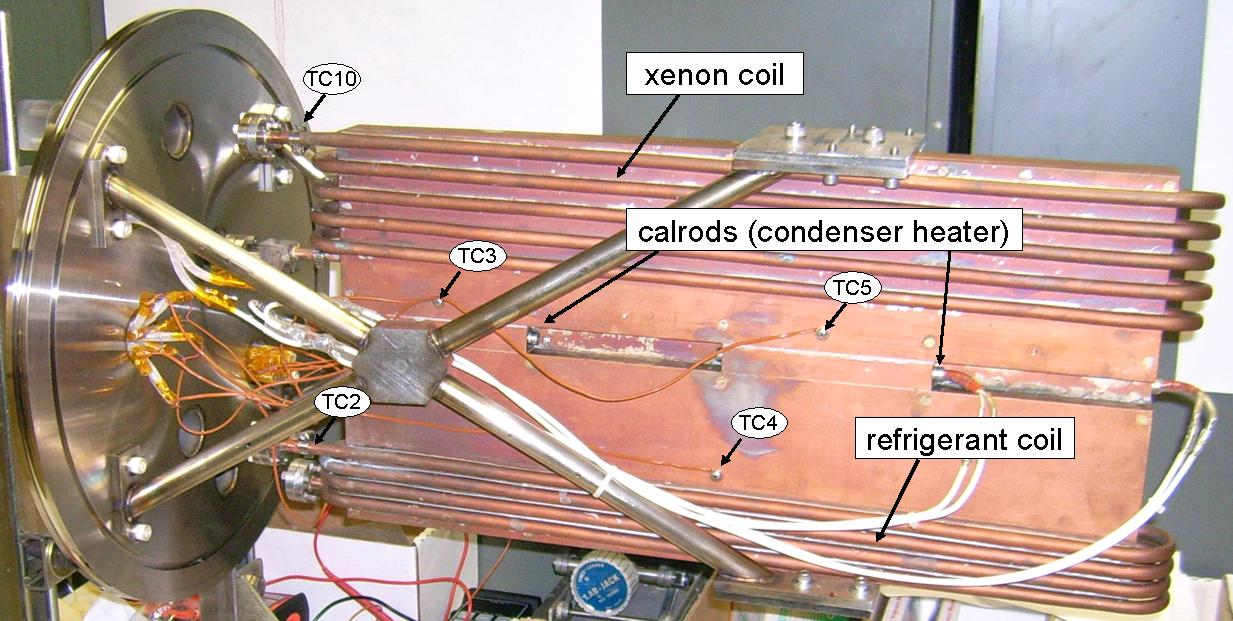}
	\caption{The LXe condenser removed from its insulating vacuum can.\label{fig:lxecondenser}}
\end{figure}

The condenser temperature was tightly controlled since the condensation rate, and therefore the Xe system pressure, depended critically on it.  This was complicated by the fact that the condenser temperature resulted from a balance of large heating and cooling powers.   Heat leaks amounted to \SI{250}{W}, while heating from the cooling and condensing of the incoming Xe gas ranged from \SI{0}{W} with no Xe recirculation to around \SI{200}{W} at \SI{20}{SLPM}.  The refrigerators used to cool the condenser were similar to those used to cool the cryostat (section \ref{sec:fridges}), but were adjusted by the manufacturers to better match the condenser heat load.  A Polycold 552HC originally provided about \SI{600}{W} of cooling, but it was replaced after developing internal refrigerant leaks.  A Telemark TVP2000 using a special Ar-enriched refrigerant mixture was used for the bulk of detector operation and provided about \SI{780}{W} of cooling.  The difference between these heating and cooling powers was made up by the trim heater.  The response of the trim heater control was therefore required to be very nimble, and selecting the correct PID parameters was challenging.  In practice, temperature variations in the control TC were typically no more than \SI{0.05}{K}.  

A heater consisting of a resistive heating element inserted in a copper cylinder provided the energy to vaporize LXe during recirculation.  LXe travelled through the cylinder, absorbing the heat added to it.  Unlike the condenser, the xenon heater did not have a temperature setpoint.  Instead, the input to its PID loop was the TPC vessel dP.  This allowed the heater to follow naturally the manual adjustment of the xenon recirculation speed and condenser temperature during recirculation, producing more or less vapor as needed to keep the pressure stable. This arrangement also provided fine dP control within the feed/bleed system deadband (see section \ref{pcon}) during steady state operation.

The Xe condenser and heater connected to the TPC through vacuum-insulated transfer lines designated the ``supply" and ``return'' lines respectively.  These transfer lines extended between two of the modules that made up the cleanroom (see section \ref{sec:EI}), and therefore incorporated flexible bellows to allow for some relative movement between those modules.  In normal operation these transfer lines, as well as an additional vacuum-insulated "recovery line" connecting the heater to the gas-handling system, contained liquid Xe.  The heights of these two LXe columns were measured using differential pressure sensors identical to those used for the redundant dP measurement (section \ref{pcon}).  Filling the supply line required adjustment of the condenser temperature to liquefy enough xenon to offset vaporization produced by heat leaks.  

A gas phase xenon pump \cite{pump}, specially-designed to maintain the xenon purity and using no lubricants, circulated xenon gas from the heater through the purifiers to the condenser.  The flow rate varied significantly over the course of each pump stroke, so the average recirculation rate was set lower to avoid exceeding the \SI{20}{SLPM} maximum rate, at about \SI{14}{SLPM} in Phase I and \SI{16}{SLPM} in Phase II.  The wearing of gaskets in the xenon pump caused the average recirculation rate to slowly diminish over time.  The pump speed was increased approximately monthly to maintain the desired average flow to within about \SI{2}{SLPM}.  The gaskets were replaced approximately annually as wear exceeded what could be compensated for by such increases in pump speed.

The purifiers were operated at an elevated temperature of \SI{550}{\degreeCelsius} in order to efficiently remove the long chain hydrocarbon and heavy poly-fluorinated contaminant, apparently residue from the isotopic enrichment process, found during the first fill with enriched xenon \cite{Auger:2012gs}.  External PID-regulated heater controllers were used to maintain this temperature, which was beyond the factory-supplied setpoints for the devices.  The decision to increase the operating temperature of the purifiers was made after consultation with SAES experts.

Both purifier cartridges were found to leak after the xenon was recovered to the storage cylinders at the end of Phase I.  During the period when the system was filled with xenon gas only, one of these leaks either developed or increased enough to lower the system pressure to atmospheric pressure in a few hours.  Limitations on underground access at WIPP at that time (see section  \ref{events}) meant that the purifiers remained heated without xenon flow for about six months prior to this.  A leak of this magnitude would likely have resulted in the loss of significant quantities of enriched xenon had it occurred while the TPC was filled with LXe.  The leaks in both cartridges were through ceramic feedthroughs used for cartridge lifetime monitors.  It could be that the leaks were related to the operation of the cartridges at elevated temperatures and/or without recirculation flow.  Since EXO-200 did not use these factory-supplied lifetime monitors, custom cartridges without the corresponding feedthroughs were obtained from SAES for Phase II. 

Establishing recirculation required careful coordination of the evaporation at the heater, gas flow through the purifiers, and recondensation at the condenser.  This was accomplished with iterative adjustment of the pump speed and condenser temperature setpoint, with the slow control adjusting the heater power to control the TPC vessel dP.  Power outages upset this balance by shutting down the condenser refrigerator, so an interlock was developed in the slow control to stop the recirculation pump and heater power to limit the resulting pressure instabilities.

\subsection{Transfer}
\label{transfer}

To fill the detector, the cryostat temperature and Xe system pressure were adjusted so that gas condensed on the inner surfaces of the TPC vessel. The feed system countered the resulting drop in pressure by adding more gas to the system, which was in turn condensed, and this process was allowed to continue until the TPC was full.  Once full, the TPC vessel's condensing power decreased substantially, since only the relatively small surface of liquid at the input plumbing remained as a condensing surface.  The external condenser (section \ref{sec:recirc}) was used to complete filling of the external LXe plumbing to levels used during normal operation.  

To empty the detector, the Xe heater (section \ref{sec:recirc}) was used to vaporize LXe, and the resulting gas was removed by the bleed system to counteract the corresponding pressure increase.  A small amount of LXe in the return line between the TPC vessel and the Xe heater was left behind after this process had emptied the rest, and that remainder could be removed either gradually by warming the entire cryostat or quickly by venting the vacuum insulation in the return line.  About \SI{36}{h} was required to recover the bulk of the Xe using the heater, with the rate of recovery limited by the capacity of the bleed system to remove the Xe gas.  Once the system pressure was lowered below the minimum compressor input (see section \ref{pcon}), a cryopump bottle connected to the buffer volume was used to recover The Xe remaining in the system below the minimum compressor input pressure (see section \ref{pcon}) was cryopumped into a 

Provision was also made for recovering the Xe on an emergency basis using pre-programmed settings for the Xe heater and HFE pressure control.  This "auto recovery" could be invoked by pressing a button on the control panel, or the system could independently begin the recovery process in the event that electrical power and cryostat cooling were lost.  In that scheme the latter conditions would be taken as evidence that the UPS (see section \ref{sec:EP}) powering the cryostat refrigerator had failed and that the remaining UPS should be used to power Xe recovery.  The operator could intervene to either pause or accelerate the process, but the process was designed to begin recovery under the above conditions if the operator was unable to intervene.  Auto recovery was never used to recover any significant quantity of Xe.

The filling and emptying processes were carried out a total of five times over the life of the experiment, including three times with the enriched Xe stockpile.  While control strategies evolved over these operations, the feed and bleed systems generally performed as expected.  A few exceptions are described below.  

In order to achieve stable feed pressure during filling, it was found necessary to actively heat the regulators to counteract the cooling effect of the expanding feed gas. This heating was applied with external heaters installed on plumbing immediately upstream of the regulators and was not required for the less frequent feeding typical of other detector operations.   In addition, it was found necessary to manually adjust the feed regulator settings when filling or emptying the detector in order to account for the supply pressure effect.  The manufacturer specification for the regulator was a change in output pressure of 3.5\% of the change in supply pressure, which would result in a \SI{0.18}{MPa} change to the \SI{0.21}{MPa} feed regulator output setting.   Since such manual adjustment was not possible without access to the WIPP underground, the configuration adopted for Phase II data collection had one feed regulator adjusted for the low xenon supply pressure occurring when the detector was full and the other adjusted for the higher supply pressure occurring when the detector was empty.  This arrangement ensured feed capability for pressure control in both scenarios at the cost of feed circuit redundancy.

The bleed system compressors were found early on to require long times (hours) to prime, inconsistent with the brief, intermittent pumping needed to maintain the buffer volume pressure during bleeding.  These times were reduced to minutes or seconds after pumps were added to the oil circuit plumbing in both units.  In addition, in at least one instance, a leak in one of the compressor heads was observed, so that the compressors were suspected of causing the air contamination found in some of the natural Xe \cite{cold_trap}.  While subsequent investigation of the head was inconclusive, the leak was presumably around the diaphragm o-ring seal and possibly could have been avoided by the use of all-metal-sealed compressors \cite{lz}.

\subsection{Xe Purity}
\label{purity}

Removal of select volatile electronegative impurities contaminating the xenon, and minimizing the sources of these impurities, were driving concerns for the xenon system. There were several possible sources in the system.  First, impurities were inherent in the bottled xenon at the ppm level, though the variation between bottles was significant \cite{cold_trap}.  Second, the construction of the high pressure portion of the system included components not rated for ultra-high-vacuum, raising the possibility of small air leaks.  Finally, all components in the xenon system, including the plumbing itself, outgassed impurities at some level.  Prior to operation, all connections in the xenon system outside of the cryostat insulating vacuum were helium leak-checked using a residual gas analyzer (RGA) connected to the system.

In steady state operation of the xenon system, the pressure in the LXe vessel was regulated by the Xe heater, allowing the block valves of the feed/bleed system to remain closed almost indefinitely (section \ref{pcon}). This essentially eliminated the first two sources of impurities described above.  Nevertheless, the purity was drastically reduced on those occasions when xenon was fed into the system \cite{prc}.   Typically this occurred when a power outage (see section \ref{sec:EP}) shut down the Xe condenser refrigerator.  In this case the approximately 0.4 kg of LXe in the supply line was bled away to protect the TPC while the condenser warmed, and was then fed back in to refill the supply line after the power was restored. The purity typically required a few days to recover afterward.  These reductions in purity and the slight resulting loss of physics data could have been avoided if sufficient UPS battery power (see section \ref{sec:EP}) had been available to support the condenser refrigerator during extended outages.  For planned outages of less than 8 hours in Phase II, however, the condenser refrigerator was successfully powered on UPS, preserving the Xe purity at the expense of not powering either cryostat refrigerator.  A different approach would have been to establish a small source of pre-purified feed Xe to use in such instances.

The TPC itself was considered the most concerning potential source of outgassing, given the high solubility of chemical impurities in the large amount of Teflon, acrylic, and kapton present inside and therefore in direct contact with the LXe.  To address this concern, those components underwent an inert gas purge before installation in the TPC \cite{Auger:2012gs}, following laboratory R\&D evidence that such treatment would reduce outgassing in LXe.  As an additional measure, almost two months of Xe gas recirculation through the purifiers to remove outgassed impurities were carried out after the TPC was installed in the cryostat.  This was done before cooling down the detector, since more outgassing was expected at the higher temperature.   

Two methods were employed to evaluate the purity of the gaseous xenon.  The relative purity was continuously monitored by Gas Purity Monitors (GPMs) \cite{gpm}, while absolute concentrations of select electronegative impurities and radioactive backgrounds were determined from occasional samples by cold-trap-enhanced mass spectrometry \cite{cold_trap}.  The GPMs were in use for the majority of Phase I data collection, and were particularly crucial to the detection of the contaminant in the enriched xenon (see section \ref{sec:recirc}) during the first TPC fill with that stockpile.  The GPMs were not used in Phase II, when they were found to read low currents uncorrelated with the xenon purity as measured in the TPC.  The reason for this failure is not understood. 

\begin{figure}
	\begin{center}
        	 \includegraphics[clip=true,trim=1.5cm 2cm 1cm 2cm,width=5in]{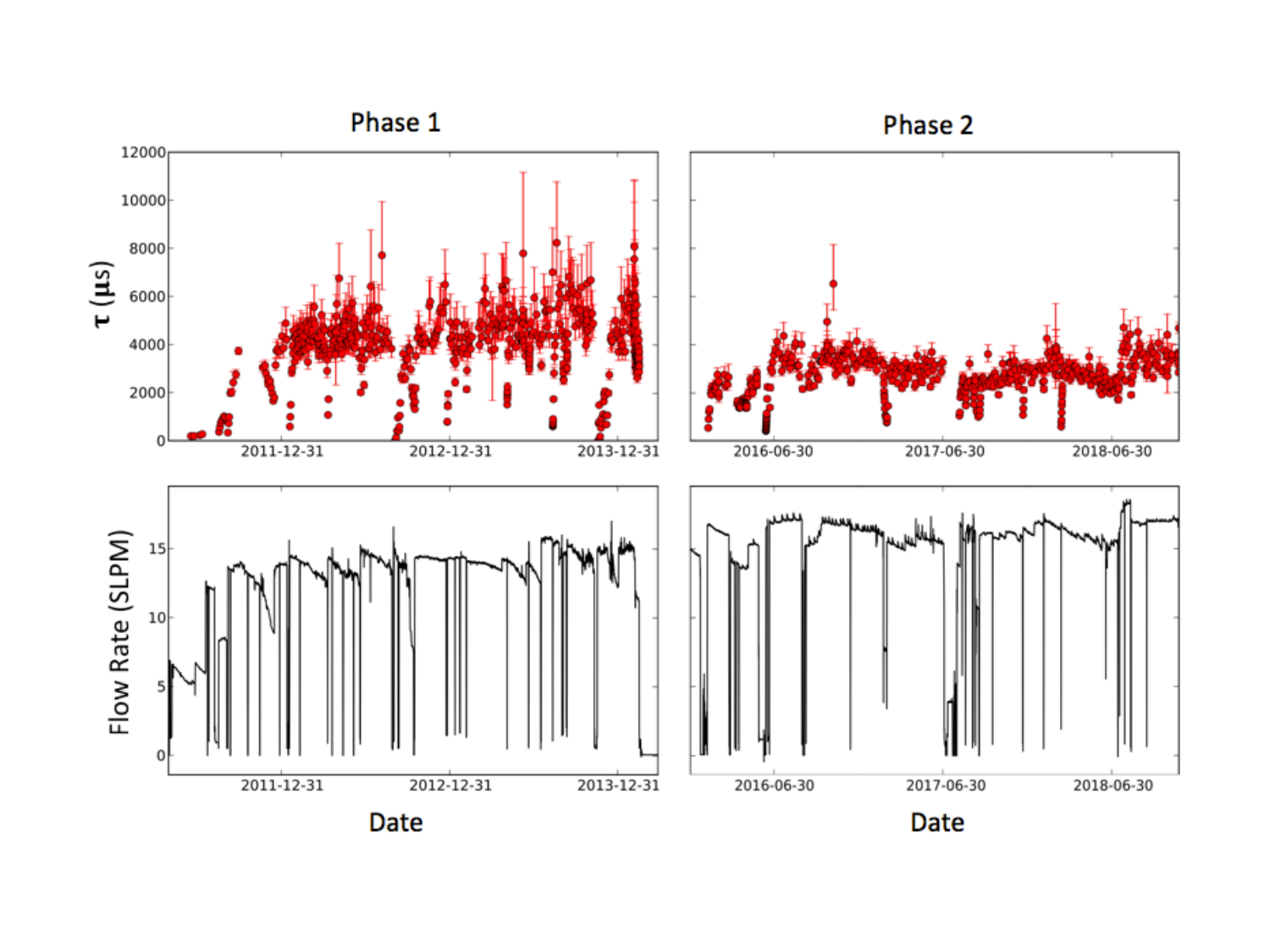}
	\caption{Electron lifetime (top), measured with the standard $^{228}$Th source calibrations discussed in the text, and xenon recirculation rate (bottom) vs. time for Phase I (left) and II (right).  These measurements were not available for the first 6 weeks of Phase I.  Each phase begins with a detector fill, with a third fill carried out in early July 2017.  Interruptions in recirculation in both phases generally corresponded to power outages.\label{fig:purity_plot}}
	\end{center}
\end{figure}

The purity of the liquid xenon in the TPC itself was determined from the attenuation of charge collected from gamma-ray events in calibration runs with radioactive sources.  A standard calibration with a $^{228}$Th source deployed to the ``S5'' position near the cathode was carried out every few days  \cite{prc}, and the resulting electron lifetimes are shown in Figure \ref{fig:purity_plot}.   Equilibrium electron lifetimes between 2 and 5 ms were measured at average recirculation flow rates between 14 and \SI{16}{SLPM}, consistent with the \SI{1}{ms} design goal (see section \ref{intro}), while about \SI{250}{\micro\second} was obtained with the lower flow rate of around \SI{6}{SLPM} in the beginning of Phase I. Equilibrium lifetime was reached after 1-3 weeks of recirculation following detector filling and generally tracked the recirculation rate, though the value for a given recirculation rate was systematically lower in Phase II than in Phase I.  The reasons for this difference are not completely understood and possibly relate to the replacement of the SAES purifier cartridges between these periods.  The increase of steady-state electron lifetime with recirculation rate was seen to be faster than linear when measured in special tests over a large range from 4 to \SI{18}{SLPM}, and this dependence will be discussed further in the upcoming final detector performance paper.

\subsection{Radon}
\label{sec:radon}

The presence of $^{222}$Rn from the uranium series in the LXe was an expected background for the experiment, since one of its decay daughters ($^{214}$Bi) emits a gamma ray near the $^{136}$Xe $\znbb$ Q value.  The emanation of $^{222}$Rn was measured for most components of the xenon system as part of a screening process.  Emanation from TIG welds was reduced by using ceriated rather than thoriated electrodes.  The installation of a trap in the xenon system to address at least the  $^{222}$Rn coming from outside the TPC itself was considered as option should initial data have shown its usefulness.  Analysis of physics data, however, revealed that the background contributed to the $\znbb$ region of interest by the 200 steady-state $^{222}$Rn atoms in the active LXe was small compared to external $^{238}$U and $^{232}$Th contributions \cite{Background_PRC}, and therefore no trap was installed.  The Rn level was, however, elevated following Xe feed events, particularly after power outages and the associated refilling of the Xe supply line (see section \ref{purity}) with Xe from the storage bottles.  The Rn level then decayed with the expected \SI{3.8}{d} half-life, typically reaching a level consistent with data-quality cuts \cite{prc} during the few days needed for purity recovery and so not requiring additional physics data to be cut from analysis.

\section{HFE system}
\label{sec:HFE}

\begin{figure}
	\begin{center}
       \includegraphics[width=5in]{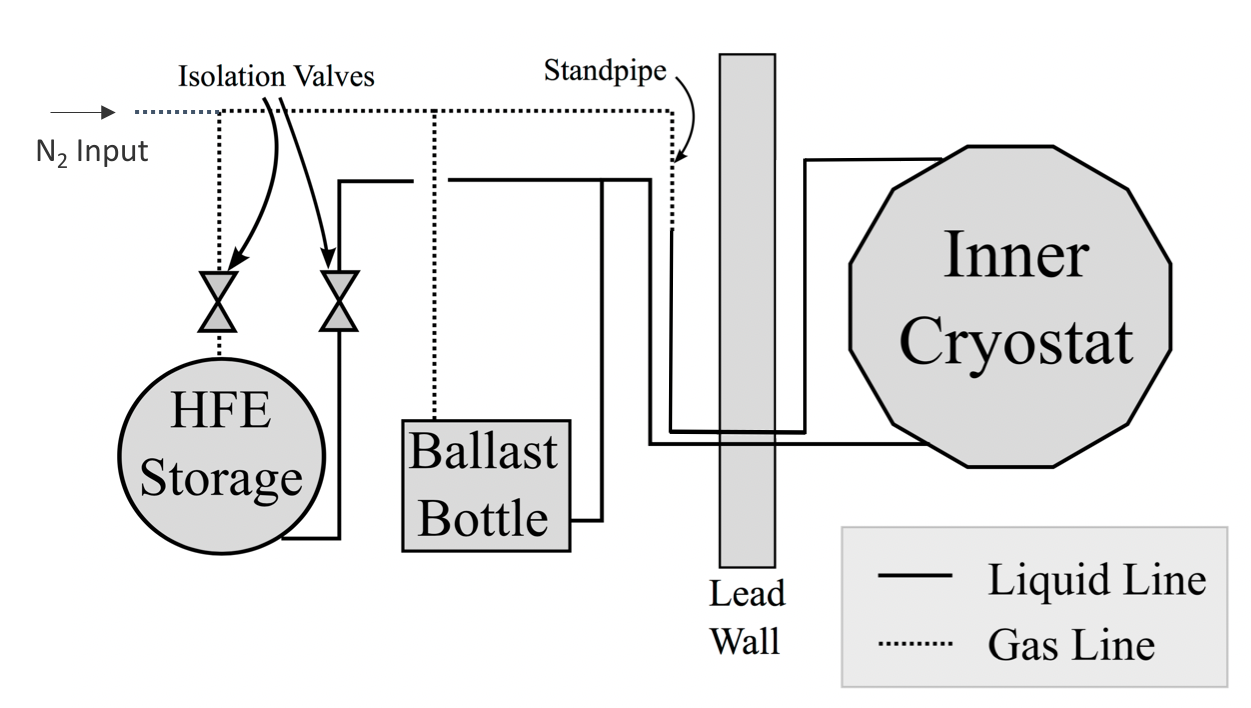}
        \caption{A simplified schematic of the HFE system.  The major system components are the storage dewar, ballast bottle, and inner cryostat vessel (IV).\label{fig:HFEsystem}}
        \end{center}
\end{figure}

A simplified schematic of the HFE system is shown in Figure \ref{fig:HFEsystem}. Its primary functions were transfer of HFE to and from the IV and control of the external pressure on the TPC vessel immersed in the HFE.  The latter effectively set the pressure in the xenon system as well, since the xenon pressure was adjusted to regulate the TPC vessel dP.

The HFE was stored in a custom cryogenic dewar supplied by Wessington Cryogenics \cite{wessington}.  The dewar had two ports, one with a dip tube that reached the bottom, and a second which accessed the vapor above the liquid.  HFE was transferred from the dewar to the cryostat and ballast bottle (see below) by pneumatic pressure applied on top of the liquid surface through the vapor port, which pushed the HFE up through the liquid port and into the rest of the system.  The pneumatic pressure was supplied with ultra-high purity (>99.999\%) N$_{2}$.  A mechanical pump to drive HFE transfer was included in the system, but was found too difficult to keep primed during operation, and the pneumatic procedure was adopted instead.   

HFE liquid density increases from 1.40 g/cm$^3$ at room temperature to 1.77 g/cm$^3$ at 170K (see Table \ref{tab:HFEprop}). About 2500 L (3500 kg) of HFE was required to fill the cryostat at room temperature, and an additional 900 kg was added during cryostat cool down.  This was handled automatically by computer-controlled valves between the cryostat and the storage dewar, feeding back on the height of liquid HFE in the vertical standpipe connected to the cryostat.  Another $\sim$100 liters (\SI{140}{kg}) filled HFE plumbing outside of the cryostat.  

The HFE pressure process variable (PPV) was defined to be the pressure in the IV at the TPC midplane.  This was calculated from pressure measurements either at the inlet port at the base of the OV or at the top of the standpipe and included a correction to get the correct liquid head at the midplane.  Since no pressure measurement was available at the OV outlet port, the head correction could not be derived from the pressure difference between the IV top and bottom, in analogy with the Xe PPV (section \ref{pcon}).  Instead, the correction was derived from the measured mass of HFE transferred from the storage dewar to the IV.  The HFE PPV was controlled to within $\pm$\SI{0.33}[]{kPa} of a \SI{138.6}{kPa} setpoint during normal operation. 

The HFE PPV was controlled using a ``ballast bottle'', a 120 L stainless steel tank filled approximately halfway with liquid HFE.  Like the storage dewar, the ballast bottle had one port for the vapor phase on top and another reaching down to the liquid, and these ports were connected to the vapor and liquid plumbing, respectively, between the storage dewar and the cryostat.  Two 200 \si{W} Peltier thermoelectric heater/coolers (TECs) \cite{tec} were mounted on the outside of the ballast bottle near the liquid surface.  The pressure in the vapor between the ballast bottle and standpipe liquid HFE, and therefore in the cryostat liquid HFE, could be increased or decreased by heating or cooling, respectively, the ballast bottle liquid HFE.  Heating generally maintained the HFE pressure set point, and the cooling capability of the ballast bottle was intended to handle temperature increases in the cleanroom stemming from, for example, power outages.  The liquid level in the ballast bottle was determined from the difference in pressure measurements at the top and bottom and the density of HFE.  All pressure measurements in the HFE system, including those used for the HFE PPV, were made with Omega PX305 transducers \cite{omega}.  

The ballast bottle was thermally insulated with foam on the outside.  The TECs were mounted with bandclamps on sheets of thermally conductive foam to help improve thermal contact with the HFE across the stainless steel tank wall.  The vertical vapor plumbing on top was in the form of a spiral of copper tubing to provide flexibility for assembly while also eliminating local minima (``U'' bends), which were found to collect liquid HFE and complicate the pressure relationship between the ballast bottle and standpipe vapor spaces.  Power to the Peltier heater/coolers was supplied by 60V/20A DC power supplies operated by a PID loop in the slow control software (see section \ref{sec:SC}) to keep the measured HFE system pressure at the programmed setpoint.  Relays changed the polarity of the supplied voltage between heating and cooling.  For robustness, there were two identical power supplies, each connected to an independent TEC and controlled by a different slow control PLC.

The mechanism for this pressure control scheme was intended to be the change in HFE vapor pressure as a function of temperature.  In practice, however, it was found that an admixture of N$_{2}$ in the ballast bottle was necessary for the control to be effective, perhaps indicating that the dominant mechanism was the change in solubility of N$_{2}$ in the ballast bottle HFE as a function of temperature.  This possibility was supported by an observed trend during Phase I in which increasing TEC cooling power, rather than constant heating, was needed to hold the HFE pressure setpoint over a period of 18 months.  Since occasional venting of the ballast bottle vapor plumbing to the outside drift was found to counteract this trend,  the effect may have stemmed from N$_{2}$ dissolved during HFE transfer coming out of solution. No such trend was observed in Phase II.  The difference may have corresponded to the fact that Phase I data collection began with pneumatic transfer of cold HFE, whereas Phase II began with a transfer at room temperature, since the solubility of N$_{2}$ in HFE is enhanced at lower temperatures.  While detailed calculations of the quantity of dissolved N$_{2}$ were not attempted, information provided by 3M indicated that approximately equal volumes of dissolved STP N$_{2}$ would be present in the HFE on delivery.

Pressure protection for the IV (section \ref{sec:cryostat}) was provided through relief valves.  Two valves, one to the vapor port on the storage dewar and the other to atmosphere, were operated by the slow control with programmable limits generally set at \SI{180}{kPa} and \SI{200}{kPa}, respectively.  A mechanical relief valve to atmosphere at \SI{207}{kPa} was also included.

\section{Slow Control}
\label{sec:SC}

\begin{figure}[!]
\begin{center}
\includegraphics[width=1.0\textwidth]{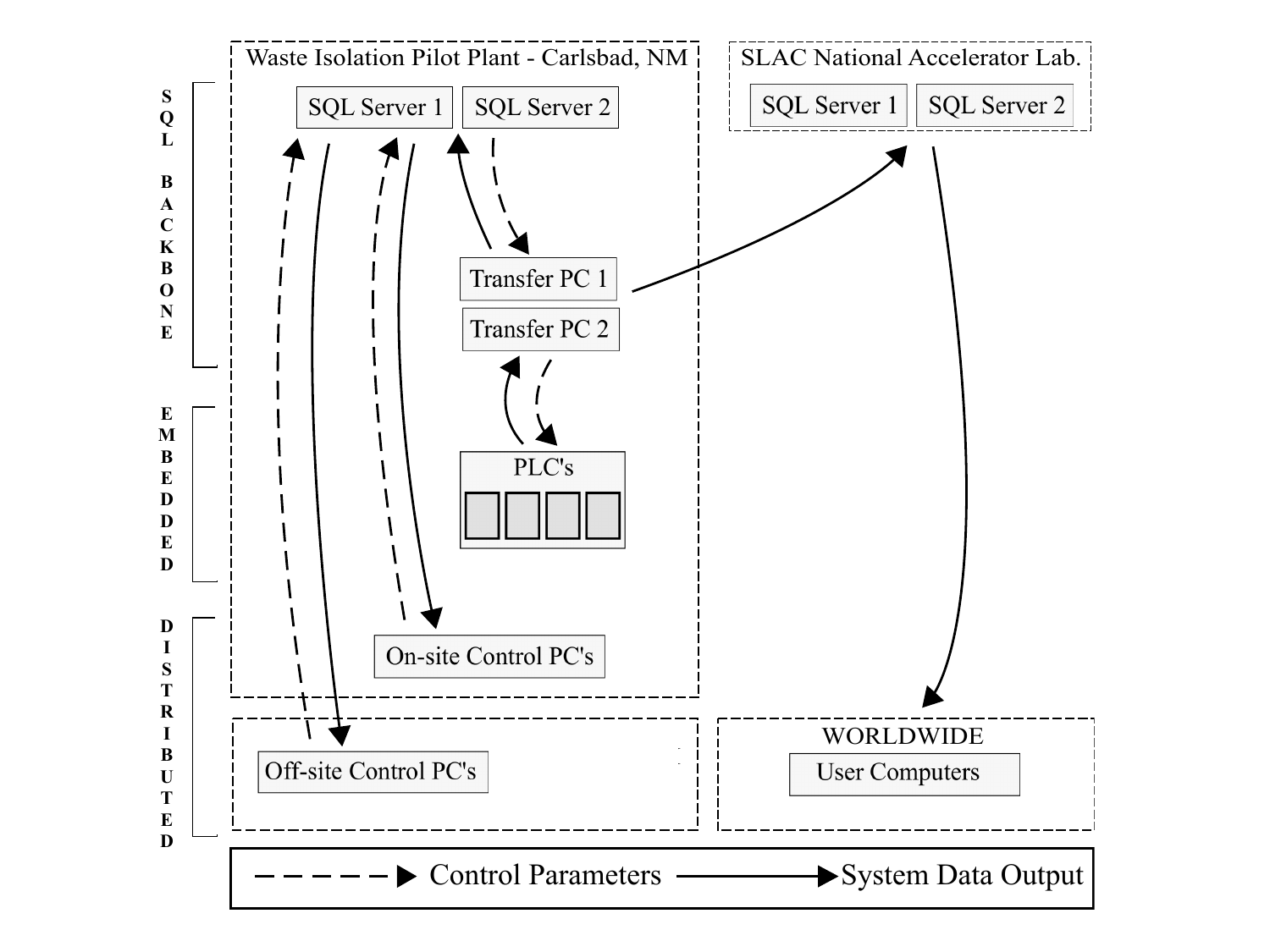}
\end{center}
\caption{The EXO-200 slow control architecture consisted of three major parts- the MySQL backbone, the embedded system, and the distributed system.}
\label{fig:arch}
\end{figure}

The slow control system (Figure \ref{fig:arch}) consisted of a network of embedded National Instruments Compact FieldPoint \cite{cfp} nodes communicating with a distributed user interface through a MySQL \cite{mysql} backbone.  The nodes included four programmable logic controllers (PLCs) and three ethernet extensions.  Once every second, LabVIEW programs running on the PLCs read in system data (typically voltages and currents) and issued control decisions to system components (heater power supplies, valve controllers, etc) through I/O modules mounted on the nodes.

Approximately 700 data channels were assembled and could be viewed in real time for system monitoring.   System data were also stored for later analysis using the open source MySQL Relational Database Management System, selected for its proven stability, flexibility, and scalability in diverse applications.  There were two dedicated database servers underground at WIPP and another two, in a master-master system, at the SLAC National Accelerator Center in Menlo Park, California.  The system dataset was published over the experiment's computer network and written by transfer programs both to the local and remote MySQL servers.  The two databases at WIPP also each held a table of control parameters for the system (thresholds, set points, etc), which were written by the transfer programs to the PLCs.  The transfer programs each read and wrote data once every second.

The data were originally transferred to and from the PLCs via LabVIEW network shared variables, each hosted on one node and accessed by the others.  This arrangement required the transfer programs to run on Windows PCs in order to support the necessary National Instruments software and was also found to prevent the PLCs from being restarted independently, which in turn complicated the process of updating the LabVIEW code.  To improve maintainability, the system was later modified to transfer the data through simple network protocols, which also allowed the transfer programs to be ported to Python running on the database servers.
  
\begin{figure}[!]
\begin{center}
\frame{\includegraphics[width=0.9\textwidth]{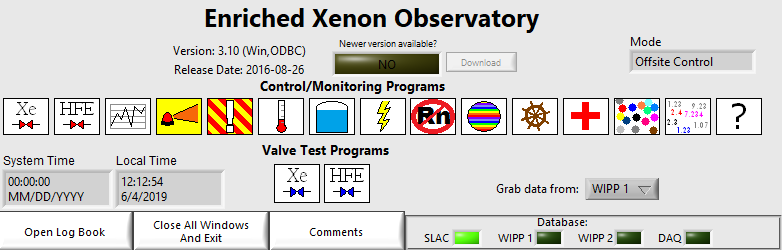}}
\end{center}
\caption{EXO-200 Distributed Software Front Page.}
\label{fig:front_page}
\end{figure}

A piece of distributed software (Figure \ref{fig:front_page}) running as a standalone executable allowed users to interact with the data.  Computers on site or at dedicated offsite control centers could access the database servers underground, while others were limited to the SLAC servers in order to preserve the limited bandwith available underground at WIPP for collaborators actively on shift monitoring the system.  Access to the WIPP servers allowed users to control EXO-200 by changing parameters in the control tables.  This required a username and password and any changes were logged in a MySQL table for accountability and reference purposes.

A number of possible alarm conditions were calculated by the embedded system and corresponding alarm data channels were added to the dataset.  The value of these alarm channels was either zero, indicating no alarm condition, or an error code. The distributed software assembled these alarm channels and displayed to the user a list of active alarms along with an auditory signal. The user had the option of bypassing alarms for a certain amount of time. User interaction with alarms was logged for accountability and reference purposes. 

The slow control was critical to detector operation, including protection of the thin-walled TPC and the enriched xenon it contained, and careful thought was given to its robustness.  Each part of the system incorporated redundancy to prevent single-point failures. The MySQL backbone had completely redundant servers and transfer programs, and control parameters could be written to the databases from any control-enabled computer.  For critical input data, such as pressure measurements used to determine the differential pressure across the TPC vessel, there were two redundant sensors, each read by a different PLC and associated I/O modules.  Critical output components, such as the HFE ballast bottle TECs, were similarly redundant and divided between the PLCs.  The embedded system could continue to operate independently of contact with the transfer programs or databases by using the last set of control parameters received.  The PLCs also checked new values of the control parameters against predefined validity ranges before accepting them, in case bad values were written to the database or the data were corrupted in transmission.
 
 A subset of the embedded system consisting of two PLCs (MPR1 and MPR2) and two ethernet extensions (MET1and MET2) read or wrote critical system data and were each powered by redundant \SI{24}{VDC} supplies which in turn were each powered from a different UPS (A or B, see section \ref{sec:EP}).  Normally MPR1 made system decisions and output commands to both its local hardware and hardware on the other nodes.   If communication between the PLCs was lost, MPR2 would operate its local hardware independently.  It was originally intended that each PLC be able to operate the detector with only the hardware located on its own backplane, but several data channels read on two of the ethernet extensions were also found to be critical, including the cryostat temperature and other parameters used by the emergency mode logic (see section \ref{sec:xetransfer}).  Loss of network communication between these four nodes would have made cryostat temperature control impossible and would have triggered automatic recovery of the xenon, so a robust industrial network switch with redundant power supplies was installed.

Failure of the PLCs was originally thought to be unlikely, since the mean time between failure (MTBF) of the devices used is quoted to be 34.4 years at \SI{25}{\celsius} and somewhat longer at the nominal clean room temperature of \SI{21}{\celsius}.   Nevertheless, execution of the LabVIEW code stopped on a PLC a few times in nearly eight years of operation for unknown reasons.  In these events, the redundancy between the PLCs discussed above allowed crucial functions including detector pressure maintenance to continue, though xenon recirculation was negatively affected.

\section{Electrical power}
\label{sec:EP}

The loss of electrical power to the auxiliary systems would have created a dangerous scenario in which the LXe warmed without active cooling and the bleed system could not have regulated the resulting increase in gas pressure.  To avoid such a scenario, the experiment had two levels of protection against power loss.  First, the experiment was powered from two separate, redundant feeds from the surface through an automatic transfer switch (ATS).  Second, critical systems were powered by two large (30 kVA) uninterruptible power sources \cite{apc} (UPSA and UPSB).  These units and their associated battery banks were located in two dedicated ISO containers in the drift outside the cleanroom laboratory (see section \ref{sec:EI} below).

Individual power feeds were lost several times a year, while the simultaneous loss of both feeds typically occurred once or twice a year.  These outages had various causes, ranging from planned maintenance onsite to weather-related impact on the local power grid.  For most of detector operation, the ATS was programmed to switch between feeds within \SI{1}{\second}, and such switches typically did not impact detector operation beyond the occasional nuisance equipment trip.  

Strategically, UPSA powered the refrigerators to maintain the cryostat temperature for \SI{24}{\hour}, while UPSB powered the Xe compressors to recover the Xe back to storage bottles after UPSA was exhausted.  Control systems and redundant sensors were powered from both.  These units successfully powered the experiment through outages of both feeds lasting up to \SI{17}{\hour}, well within the expected UPSA runtime.  It never became necessary to recover the xenon to storage on UPSB battery power.

Each UPS utilized a bank of 128 \SI{12}{V} batteries, arranged in four parallel strings of 32 connected in series, for a total capacity of \SI{220}{kWh}.  The use of such large battery systems is unusual.  Alternative methods of providing backup power were considered, but were rejected in negotiations with WIPP in view of the unique constraints related to being underground.  For example, the use of a diesel backup generator underground was not permitted because of concerns over carbon monoxide accumulation.  Both UPS units were completely rebuilt after the 2014 WIPP fire (see section \ref{events}) because of concerns about soot contamination.  The batteries were replaced twice, once at the end of their normal life and again after they were completely drained in the aftermath of the 2014 WIPP events.
 
UPSB power was lost twice.  On the first occasion the unit failed during a power outage, and its loads remained unpowered for about \SI{4}{\hour} until personnel could be sent underground to manually bypass the unit.  Though the resulting unanticipated combination of system failures caused a large TPC vessel dP of \SI{142}{kPa}, well beyond the \SI{35}{kPa} rating, the experiment was not damaged.  On the second occasion, a single output phase was lost, disabling some but not all of the UPSB loads.  These failures are not understood and may represent an inherent risk in using UPS backup power.  

In addition to underground power outages, the experiment could be negatively affected by the loss of power to the EXO office in the WIPP 
Support Building, where network equipment necessary for communication with the offsite control centers (section \ref{sec:SC}) was housed.  A combination of rack-mount UPS units and a small (\SI{5.5}{kW}) gasoline-powered generator operated just outside the building provided power to maintain this communication through infrequent outages.

\section{Experimental installation}
\label{sec:EI}

The EXO-200 experiment was located \SI{655}{\meter} underground in the North Experimental Area (NEXA) at the WIPP site, a Department of Energy facility that disposes of defense-generated transuranic waste in an ancient salt formation.  This disposal strategy takes advantage of bedded salt's tendency to seal and close cavities.  The atmospheric pressure underground is about \SI{96}{kPA}.

The main laboratory was housed in a cleanroom \cite{cleanroomswest} assembled from individual modules small enough to be brought underground by the facility's largest hoist.  Mine air passed through 3 levels of prefilters in the HVAC air handler before being sent to the main HEPA units.  Airflow to the cleanroom was arranged so that there was a progressively larger overpressure with respect to the mine for modules closer to the cryostat.   Associated infrastructure including an office, machine shops, and storage were housed in ISO containers arranged nearby.

All the cleanroom modules were supported on adjustable mounts in order to account for the movement of the salt floor.  Module 1, which contained the cryostat (6 tons) and its associated shielding (a combined 67 tons of lead and HFE), was particularly heavy and had a special reinforced floor and used three hydraulic rams for level adjustment.  The remaining cleanroom modules had four screw jacks each as mechanically adjustable support feet.  The relative level of Module 1 and the adjacent Module 2 received particular attention and frequent adjustment, since LXe plumbing to the TPC spanned that boundary.  Over time, drift closure of about \SI{8}{cm} per year resulted in close clearances for the veto panels on top of Module 1, so that salt was excavated from underneath its supports to provide additional adjustment.  The level of equipment rigidly attached to the cleanroom, such as the HVAC and the Wessington container, had to be adjusted in concert, which required retrofitting adjustable supports.   

The experiment generated approximately \SI{50}{kW} of waste heat in Phase I and \SI{80}{kW} in Phase II.  This heat was dissipated into the air, mainly by the cleanroom HVAC and the water chillers (see Section \ref{sec:TS}).  Airflow through NEXA prior to the 2014 events (see Section \ref{events}) was \SI{100000}{m^3/hr} or more, depending on the particular ventilation mode at WIPP, and the drift temperature varied seasonally between about \SI{18}{\celsius} and \SI{32}{\celsius}. After those events, however, changes in the ventilation at WIPP reduced the NEXA airflow to levels too low to reliably measure.  As a result, the average drift temperature varied slowly between \SI{30}{\celsius} and \SI{40}{\celsius} over Phase II.  While the HVAC was able to maintain normal temperatures in the cleanroom, work in the external drift was complicated by heat stress concerns.  This was particularly true for the cleanroom releveling activities and associated salt excavation.

\section{Radon-free air system}

The presence of $^{222}$Rn in the ``airgap'' between the cryostat and its surrounding lead shield was considered as a source of background.  As in the case of radon in the xenon (see section \ref{sec:radon} above), the background arises from gamma rays produced in the decay of $^{214}$Bi, a $^{222}$Rn daughter.  To mitigate this background, the cryostat and lead shielding were enclosed in a ``tent'' consisting of overlapping rubber-sealed sheet-metal segments, to be continuously purged with low radon air.  A purge system using bottled air aged to eliminate radon was installed but could not supply enough flow to overpressure the inside of the tent.  That system was replaced with a radon filter for air referred to as the ``deradonator,'' based on the device used for the Borexino cleanroom at Princeton
\cite{pocar2003}, and which will be described in a separate publication.  The deradonator began operation early in Phase II of data collection and met the design requirement of abating radon in the airgap ten-fold or more.  Analysis of Phase II data did not reveal a statistically significant reduction in the corresponding background component from external $^{238}$U \cite{phaseii_prl}, however, consistent with other evidence that airgap radon decays were not the dominant source of that background component \cite{Background_PRC}.

\section{Performance during 2014 WIPP events}
\label{events}

In February 2014 there was a fire in the WIPP underground followed 10 days later by an unrelated release from a waste drum \cite{wipp_events}.   While these events were not related to EXO-200, soot from the fire did reach the experiment.   The radioactivity release occurred in a separate mine ventilation circuit and did not directly affect the experiment.  Analysis of salt samples taken around NEXA after the events showed that no radioactive material was present.  These extraordinary events ended underground operations at WIPP for the better part of that year.  While such a long period without access to the experiment was not anticipated, the auxiliary systems were able to protect the enriched xenon, HFE, TPC, and cryostat until recovery efforts began in early 2015.

Faced with a long period without underground access, the decision was made shortly after the events to recover the xenon to the storage bottles.  This operation was carried out remotely through the slow-controls essentially in the usual way, since a xenon recovery without underground access was anticipated as a possibility during system design.  With the near-term availability of electrical power and ventilation unclear, a subsequent decision was made to begin a controlled warming of the cryostat while the auxiliary systems were still operating.  Carrying out this operation without underground access was not anticipated when the control system was designed.  Cryostat warm-up usually began by transferring the cold HFE from the cryostat to the storage dewar using pressurized N$_{2}$ as discussed above, but that required personnel underground to connect and adjust the N$_{2}$ purge.  Instead, the HFE system pressure was limited by opening a PLC-controlled valve to allow the liquid HFE to expand into the storage dewar through vapor transfer lines as the cryostat warmed.   

The warm-up took about six months.  As feared, cryostat cooling capability was lost midway during that period when the water chillers failed.  At the end of the process, while underground access was still very limited, site electrical power to the experiment was also lost and would not be restored for months until affected distribution equipment was cleaned of soot.  A brief underground entry was arranged, however, while the system was still supported by UPS power.  Since the normally-closed valve to the storage dewar would close without power, its pneumatic actuator was directly connected to a N$_{2}$ bottle and regulator to keep it open indefinitely.  Permission was not obtained from WIPP to isolate the UPS batteries underground after exhaustion, leading to their destruction.   In addition, the leak in the xenon system coinciding with the loss of power (see section \ref{sec:recirc}) brought the vessel dP to \SI{-80}{kPa}.

Without power to the slow controls, the remainder of the warm-up was carried out nearly without instrumentation.  Occasional brief underground entries were made in the months before regular access was restored in order.  A dial gauge on the storage dewar, then open to the cryostat, was the only available measure of the system pressure.  Periodic venting was required to control the pressure as the warming HFE outgassed N$_{2}$ vapor absorbed during previous pneumatic manipulations.

The process of restarting the experiment began in December 2014 as the collaboration regained semi-regular access to NEXA.  Working initially without electrical power, a fine layer of soot was vacuumed from external surfaces in the drift and maintenance needs were assessed.  As discussed in the corresponding sections above, these included extensive adjustments to the height of the cleanroom modules, replacement of the Module 1 supports, the rebuilding of both UPS units, replacement of both UPS battery banks, replacement of the cryostat refrigerators, and overdue maintenance to the HVAC and water chillers.  After replacing the leaking purifier cartridges in the xenon system and returning the HFE liquid levels to normal, the usual detector cooldown and fill procedure was completed.  Source calibrations after an upgrade to the front end electronics revealed good detector performance, and Phase II data collection with the commissioned deradonator began in April 2016.

\section{Front-end electronics upgrade}

\begin{figure}[!]
\begin{center}
\frame{\includegraphics[trim=310 540 50 100, clip, width=0.8\textwidth]{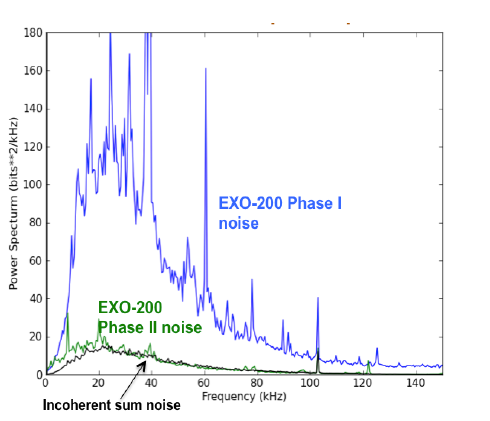}}
\end{center}
\caption{The noise power spectrum of the coherent sum of APD channels.  The Phase-I noise (blue) is reduced by a factor of 2.5 after the electronics upgrade.  The Phase-II coherent noise (green) is only 20\% higher than the incoherent sum noise (black).}
\label{fig:electronics_upgrade}
\end{figure}

Upgrades to the EXO-200 front-end readout system \cite{Auger:2012gs} were carried out before Phase-II operations to improve detector performance.  First, new front-end readout boards for APD channels were installed. These new boards used a new preamp design less sensitive to noise produced by voltage regulators.  Second, new ground adapter boards were installed to minimize ground currents between the APD channels.  Third, the shaping times for the induction wire channels were optimized to lower their signal reconstruction threshold.  As shown in Figure \ref{fig:electronics_upgrade}, the coherent sum noise of the APD channels was reduced by a factor 2.5 after the electronics upgrade.  Only 20\% of extra coherent noise remained for Phase-II data.  The excess noise was mostly in the high-frequency region outside of the sensitive frequency band of the preamp, and therefore had little contribution to the effective noise after signal reconstruction.   

Furthermore, before Phase-II operation, the cathode bias voltage of the detector was increased from -8 kV to -12 kV, changing the main drift field from $\sim$380 V/cm to $\sim$576 V/cm.  The detector operated stably at this bias voltage throughout the Phase-II operation.  The combination of lower APD noise and higher drift field improved the average detector energy resolution ($\sigma$/E) at the $\znbb$ decay Q value (2.46 MeV) from 1.25\% in Phase-I to 1.15\% in Phase-II \cite{phaseii_prl}.  In addition, elimination of the APD coherent noise lowered the scintillation reconstruction threshold, enabling the detector to probe physics channels at lower energies with the Phase-II data. 

\section{Discussion}
\label{summary}

The EXO-200 auxiliary systems enabled successful low-background detector operation with enriched xenon over a nearly nine year period.  Temperature stability and LXe purity during data collection exceeded the design goals.  The TPC vessel dP was kept within its $\pm$\SI{35}[]{kPa} design limit over a wide range of normal and off-normal conditions, with two exceptions that did not negatively affect vessel integrity or TPC performance.  Key to this success was protection from single points of failure through redundancy throughout the auxiliary systems.  Nevertheless, the experiment's operational history makes clear that reliable electrical power, personnel access, and heat dissipation are essential requirements in the operation of a complex underground cryogenic experiment.

\acknowledgments{EXO-200 is supported by DoE and NSF in the United States, NSERC in Canada, SNF in Switzerland, IBS in Korea, RFBR (18-02-00550) in Russia, DFG in Germany, and CAS and ISTCP in China.  EXO-200 data analysis and simulation uses resources of the National Energy Research Scientific Computing Center (NERSC). We gratefully acknowledge the KARMEN collaboration for supplying the cosmic-ray veto detectors, and the WIPP for their hospitality.  We thank Randy Elmore of WIPP for the initial idea of the thermal store.  We thank SDMS for their commitment during the construction of the cryostat, and Sheedy Drayage for the careful shipment of the experiment to the WIPP site.  We thank Carriaga Machine for onsite support during installation and operations. Finally, we thank H. Mong and W. Jarratt for assistance with the preparation of figures.}

\bibliographystyle{h-physrev3}
\bibliography{exo200partII-v4.0}

\end{document}